# Regulating ChatGPT and other Large Generative AI Models

Working Paper, this version May 12, 2023

Philipp Hacker

European New School of Digital Studies, European University Viadrina, hacker@europa-uni.de

Andreas Engel

Heidelberg University, andreas.engel@igw.uni-heidelberg.de

Marco Mauer

Humboldt University of Berlin, marco.mauer@hu-berlin.de

with limited input from ChatGPT (see Acknowledgments)

**Abstract:**

Large generative AI models (LGAIMs), such as ChatGPT, GPT-4 or Stable Diffusion, are rapidly transforming the way we communicate, illustrate, and create. However, AI regulation, in the EU and beyond, has primarily focused on conventional AI models, not LGAIMs. This paper will situate these new generative models in the current debate on trustworthy AI regulation, and ask how the law can be tailored to their capabilities. After laying technical foundations, the legal part of the paper proceeds in four steps, covering (1) direct regulation, (2) data protection, (3) content moderation, and (4) policy proposals. It suggests a novel terminology to capture the AI value chain in LGAIM settings by differentiating between LGAIM developers, deployers, professional and non-professional users, as well as recipients of LGAIM output. We tailor regulatory duties to these different actors along the value chain and suggest strategies to ensure that LGAIMs are trustworthy and deployed for the benefit of society at large. Rules in the AI Act and other direct regulation must match the specificities of pre-trained models. The paper argues for three layers of obligations concerning LGAIMs (minimum standards for all LGAIMs; high-risk obligations for high-risk use cases; collaborations along the AI value chain). In general, regulation should focus on concrete high-risk applications, and not the pre-trained model itself, and should include (i) obligations regarding transparency and (ii) risk management. Non-discrimination provisions (iii) may, however, apply to LGAIM developers. Lastly, (iv) the core of the DSA's content moderation rules should be expanded to cover LGAIMs. This includes notice and action mechanisms, and trusted flaggers. In all areas, regulators and lawmakers need to act fast to keep track with the dynamics of ChatGPT et al.

**Table of Contents**

# 1 INTRODUCTION

Large generative AI models (LGAIMs) are rapidly transforming the way we communicate, create, and work. Their consequences are bound to affect all sectors of society, from business development to medicine, from education to research, and from coding to entertainment and the arts. LGAIMs harbor great potential, but also carry significant risk. Today, they are relied upon by millions of users to generate human-level text (e.g., GPT-4, ChatGPT, Luminous, Bard, Bing), images (e.g., Stable Diffusion, DALL·E 2), videos (e.g., Synthesia), or audio (e.g., MusicLM), while further alternatives are already in the pipeline [1-3]. Soon, they may be part of employment tools ranking and replying to job candidates, or of hospital administration systems drafting letters to patients based on case files. Freeing up time for professionals to focus on substantive matters–for example, actual patient treatment–, such multi-modal decision engines may contribute to a more effective, and more just, allocation of resources. However, errors will be costly, and risks ranging from discrimination and privacy to disrespectful content need to be adequately addressed [4-6]. Already now, LGAIMs' unbridled capacities may be harnessed to take manipulation, fake news, and harmful speech to an entirely new level [7-11]. As a result, the debate on how (not) to regulate LGAIMs is becoming increasingly intense [12-22].

In this paper, we argue that regulation, and EU regulation in particular, is not only ill-prepared for the advent of this new generation of AI models, but also sets the wrong focus by quarreling mainly about direct regulation in the AI Act at the expense of the, arguably, more pressing content moderation concerns under the Digital Services Act (DSA). Significantly, the EU is spearheading efforts to effectively regulate AI systems, with specific instruments (AI Act, AI Liability Directive), software regulation (Product Liability Directive) and acts addressed toward platforms, yet covering AI (DSA; Digital Markets Act). Besides, technology-neutral laws, such as non-discrimination law, and also data protection law, continue to apply to AI systems. As we shall see, it may be precisely their technology-agnostic features that make them better prepared to handle the risks of LGAIMs than the technology-specific AI regulation that has been enacted or is in preparation.

AI regulation, in the EU and beyond, has primarily focused on conventional AI models, however, not on the new generation whose birth we are witnessing today. The paper will situate these new generative models in the current debate on trustworthy AI regulation, and ask what novel tools might be needed to tailor current and future law to their capabilities. Inter alia, we suggest that the terminology and obligations in the AI Act and other pertaining regulation be further differentiated to better capture the realities of the evolving AI value chain. Some of these observations also apply to traditional AI systems; however, generative models are special in so far as they create output designed for communication or speech–and thus raise important and novel questions concerning the regulation of AI-enabled communication, which we analyze through the lens of the DSA and non-discrimination law.

To do so, the paper proceeds in five steps. First, we cover technical foundations of LGAIMs, and typical scenarios of their use, to the extent that they are necessary for the ensuing legal discussion. Second, we critique the EU AI Act, which seeks to directly address risks by AI systems. The versions adopted by the Council (Art. 4a-c AI Act[1]) and the European Parliament (Art. 28-28b AI Act EP Version[2]) contain provisions to explicitly regulate LGAIMs, even if their providers are based outside of the EU [14, cf. also 23]. These proposals, however, arguably fail to fully accommodate the capacities and broad applicability of LGAIMs, particularly concerning the obligation for an encompassing risk management system covering all possible high-risk purposes (Art. 9 AI Act; Art. 28b(1)(a) AI Act EP Version) [12, pp. 6-10, 24, pp. 13, 51 et seqq.]. Precisely because LGAIMs are so versatile, detailing and mitigating every imaginable high-risk use seems both prohibitive and unnecessary. We also address more recent proposals debated in the European Parliament according to which LGAIMs should, at least generally, qualify as high-risk systems under Annex III AI Act. We ultimately reject this proposal. LGAIM risk regulation, in our view, should generally focus on deployed applications, not the pre-trained model [12, 24]. However, non-discrimination provisions may apply more broadly to the pre-trained model itself (more precisely: to its developers) to mitigate bias at its data source [see also 25].

Third, we highlight key data protection risks under the GDPR, with a particular focus on model inversion [26-28]. Fourth, we turn to content moderation [see, e.g., 29, 30, 31]. Recent experiments have shown that ChatGPT, despite innate protections [32], may be harnessed to produce hate speech campaigns at scale, including the code needed for maximum proliferation [8]. Furthermore, the speed and syntactical accuracy of LGAIMs make them the perfect tool for the mass creation of highly polished, seemingly fact-loaded, yet deeply twisted fake news [7, 17]. In combination with the factual dismantling of content moderation on platforms such as Twitter, a perfect storm is gathering for the next global election cycle. We show that the EU's prime instrument to combat harmful speech, the Digital Services Act (DSA) [33, 34], does not apply to LGAIMs, creating a dangerous regulatory loophole.

The paper argues for three layers of obligations concerning LGAIMs (minimum standards for all LGAIMs; high-risk obligations for high-risk use cases; collaborations along the AI value chain; cf. now also Art. 28 and 28b AI EP Version) and makes four specific policy proposals to ensure that LGAIMs are trustworthy and deployed for the benefit of society at large: direct regulation of LGAIM deployers and users, including (i) transparency and (ii) risk management; (iii) the application of non-discrimination provisions to LGAIM developers; and (iv) specific content moderation rules for LGAIMs. We conclude with a brief assessment concerning the vice and virtue of technology-neutral regulation.

Due to space constraints, we cannot address all social and regulatory concerns regarding LGAIMs and have to bracket, for example, questions of IP law, power dynamics [35, 36], a deeper exploration of the comparative advantages of technology-neutral and technology-specific regulation [37], or the use of LGAIMs in military contexts [38, 39].

## 2 TECHNICAL FOUNDATIONS OF LARGE GENERATIVE AI MODELS AND EXEMPLARY USAGE SCENARIOS

The AI models covered by this contribution are often referred to as 'foundation models' [40], 'large language models' (LLMs) [41] or 'large generative models' (LGAIMs–the term adopted in this article) [42]. Although the emergence of these models in recent years constitutes a significant technical advance (for foundations, see [43-47]), they harness, to a reasonable extent, existing technologies in a vastly increased scale and scope. LGAIMs are usually trained with several billion, if not hundreds of billions, parameters [47, 48], requiring large amounts of training data and computing power [49]. OpenAI's "CLIP" image classifier, for example, was built using a set of 400 million image-text pairs [50]. The 'BASIC' model even uses 6.6 billion such pairs [51]. While there are ongoing research efforts to make training language models, and, in particular, transformers, more efficient [52, 53], the energy required to train models this large has triggered concerns from a climate policy perspective [24, 54-58] (see also Part 7).

---

[1] Unless otherwise noted, all references to the AI Act are to the general approach adopted by the EU Council on Dec. 6, 2022, available under https://data.consilium.europa.eu/doc/document/ST-14954-2022-INIT/en/pdf; we have been able to incorporate policy developments until May 12, 2023.

[2] DRAFT Compromise Amendments on the Draft Report, Proposal for a regulation of the European Parliament and of the Council, Brando Benifei & Ioan-Dragoș Tudorache (May 9, 2023), https://www.europarl.europa.eu/meetdocs/2014_2019/plmrep/COMMITTEES/CJ40/DV/2023/05-11/ConsolidatedCA_IMCOLIBE_AI_ACT_EN.pdf (= AI Act EP Version).

**2.1 From Machine Learning to Large Generative AI Models**

Hence, LGAIMs "*are advanced machine learning models that are trained to generate new data, such as text, images, or audio*" (Prompt 1, see Annex H1). This "*makes them distinct from other AI models [... only] designed to make predictions or classifications*" (Prompt 2) or to fulfil other specific functions. This increased scope of application is one of the reasons for the large amount of data and compute required to train them. LGAIMs employ a variety of techniques [32] that aim at allowing them "*to find patterns and relationships in the data on its own, without being* [explicitly] *told what to look for.*

*Once the model has learned these patterns, it can generate new examples that are similar to the training data*" (Prompt 3). In simple terms, training data are represented as probability distributions. By sampling from and mixing them, the model can generate content beyond the training data set–thus something new, as some commentators put it [59, 60]. LGAIMs can often digest human text input [61, 62] and produce an output (text; image; audio; video) based on it. The vast amounts of data required imply that developers of LGAIMs must often rely on training data that is openly available on the internet, which can hardly be considered perfect from a data quality perspective [63]. The content generated by these models can, therefore, be biased, prejudiced or harmful [15, 64]. To avoid or at least mitigate this issue, model developers need to use proper curating techniques [65, 66]. OpenAI, controversially, hired a large content moderation team in Kenya [67].

**2.2 Content Moderation**

ChatGPT itself sums up the problem of having to curate its training data and moderate its output this way: "[*T*]*he models are designed to generate new content that is similar to the training data, which may include offensive or inappropriate content.* [...] *Furthermore, large generative models can generate synthetic content that is difficult to distinguish from real content, making it challenging to differentiate between real and fake information.* [… T]*he sheer volume of content generated by these models can make it difficult to manually review and moderate all of the generated content*" (Prompt 4). For as much as we know [32], and according to ChatGPT itself, the creators of ChatGPT sought to address this problem by using "*a combination of techniques to detect and remove inappropriate content. This process includes pre-moderation, where a team of human moderators review and approve content before it is made publicly available. Additionally, ChatGPT uses filtering, which involves using natural language processing and machine learning algorithms to detect and remove offensive or inappropriate content. This is done by training a machine learning model on a dataset of examples of inappropriate content, and then using this model to identify similar content in new inputs*" (Prompt 5). While we cannot perfectly verify these claims due to lack of transparency on OpenAI's side, it seems that ChatGPT relied or relies on humans that train an automatic content moderation system to prevent the output from becoming abusive [67].

**2.3 Two lead examples: clothing design and birthday invites**

Even (idealized) automated and perfect detection of abusive content would only solve half the problem, though. What remains is the danger of creating "fake news" that are hard to spot [17]. Regulation arguably needs to tackle these challenges. To better highlight them, for the discussion that follows, we will consider different scenarios of LGAIM use.

Consider the following two lead examples: in a business context, one might think of a sportswear manufacturer (e.g., adidas or Nike) that wants to use the potential of a LGAIM specifically for the design of clothing. For this purpose, adidas might use a pre-trained model provided by a developer (e.g., Stability AI), while another entity, the deployer, would fine-tune the model according to adidas' requirements (and possibly host it on a cloud platform). As a second exemplary use case, in a private setting, one could think of a young parent that uses an AI text generator to generate a funny (and suitable) invitation text for her daughter's birthday party. To do so, (s)he might consult Aleph Alpha's Luminous or ChatGPT and ask the chatbot to come up with an appropriate suggestion.

## 3 DIRECT REGULATION OF THE AI VALUE CHAIN: THE EUROPEAN AI ACT

On May 13, 2022, the French Council presidency circulated an amendment to the draft AI Act, Articles 4a-4c, on what the text calls "general-purpose AI systems" (GPAIS). This novel passage, which did not spark much debate initially, has surreptitiously come to form the nucleus of direct regulation of LGAIMs. It was fiercely contested in the European Parliament [68-70] and will be a key point of debate for the final version of the AI Act. The general approach adopted by the Council on December 6, 2022, defines GPAIS as systems "intended by the provider to perform generally applicable functions such as image and speech recognition, audio and video generation, pattern detection, question answering, translation and others; a general purpose AI system may be used in a plurality of contexts and be integrated in a plurality of other AI systems" (Article 3(1b) AI Act).

To ensure a "fair sharing of responsibilities along the AI value chain" (Recital 12c AI Act), these systems are subjected to the high-risk obligations (e.g., Article 8 to 15 AI Act) if they may be used as high-risk systems or as components thereof (Article 4b(1)(1) and 4b(2) AI Act). These duties arise once the Commission has specified, in implementing acts, how the high-risk rules should be adapted to GPAIS (Article 4b(1) AI Act). The territorial scope of the AI Act extends, inter alia, to providers placing on the market or putting into service AI systems in the EU, and also to situations where the output produced by the system is used in the Union. Hence, these rules may even apply if the provider is entirely based outside of the EU (Article 2(1) AI Act). Exceptionally, the provider is exempted from specific obligations for GPAIS providers if the provider explicitly and publicly excludes all high-risk uses of the GPAIS; however, the exemption fails if the exclusion is not made in good faith (Article 4c(1) and (2) AI Act). Nevertheless, if any provider detects or is informed about market misuse of its system, it must take all proportionate measures to stop the misuse and avoid harm (Article 4c(3) AI Act). This notice-and-action mechanism, structurally known from copyright law [71-73] and the DSA [31, 33, 74, 75], complements the novel, active production monitoring obligation (Articles 61, 4(2) AI Act), which had so far only been contained in the product liability tort law of some Member States [76] and will be partially introduced in the product liability upgrade as well (Article 6(1)© and (e) Product Liability Directive Proposal3 [24, 77, 78]).

### 3.1 Critique of the GPAIS AI Act Rules

The AI Act heroically strives to keep pace with the accelerating dynamics in the AI technology space. However, in our view, the recently introduced rules on GPAIS fail to do justice to the peculiarities of large AI models, and particularly LGAIMs, for three reasons.

---

[3] European Commission, Proposal for a Directive of the European Parliament and of the Council on Liability for Defective Products, COM(2022) 495 final.

*3.1.1 Toward a Definition of GPAIS*

First, the definition in Article 3(1b) AI Act is significantly over-inclusive. Rules on GPAIS were inspired by the surge in the release of and literature on foundation models and LGAIMs. As seen in Part 2, LGAIMs operate with large numbers of parameters, training data, and compute. While arguably not yet bordering on artificial general intelligence [14, but see 79], LGAIMs still are more versatile than the narrower deep learning systems that have dominated the third wave of AI so far [cf. 80, pp. 18 et seqq.]. Significantly, they can be deployed to solve tasks they have not been specifically trained for [47], and generally operate on a wider range of problems than traditional models do. Conceptually, their "generality" may refer to their *ability* (e.g., language versus vision, or combinations in multimodal models); *domain* of use cases (e.g., educational versus economic); breadth of *tasks* covered (e.g., summarizing versus completing text), or versatility of *output* (e.g., black and white versus multicolored image) [14]. GPAIS, in our view, must necessarily display significant generality in ability, tasks, or outputs, beyond the mere fact that they might be integrated into various use cases (which also holds true for extremely simple algorithms). The broad definition of GPAIS in the AI Act clashes with this understanding, however. According to that rule, every simple image or speech recognition system seems to qualify, irrespective of the breadth of its capabilities; rightly, this only corresponds to a minority position in the technical GPAIS literature [14, 81].

This problematic over-inclusiveness is caused by the second half-sentence of Article 3(1b) AI Act, where further specifications–use in different contexts and AI systems–are not formulated as (disjunctive) necessary conditions, but as merely possible (and then likely qualifying) examples of GPAIS ("may"). To specifically capture truly general-purpose systems, the definition would have to be revised so that the use of the system in different contexts or for substantively different AI systems are necessary, and not sufficient, conditions. Additionally, it should require that GPAIS display significant generality of ability, task, or output, in decreasing order of relevance [cf. also 14]. As a result, models that display only one set of abilities and tasks would need to have highly diverse output to qualify as GPAIS; conversely, multimodal models would generally qualify, even if they only apply to one specific task and output is not significantly variable.

*3.1.2 Risk Management for GPAIS*

Second, even such a narrower definition would not avoid other problems. Precisely because large AI models are so versatile, providers will generally not be able to avail themselves of the exception in Article 4c(1) AI Act: by excluding all high-risk uses, they would not act in good faith, as they would have to know that the system, once released, may and likely will be used for at least one high-risk application. For example, language models may be used to summarize or rate medical patient files, or student, job, credit or insurance applications (Annexes II, Section A. No. 12, 13 and III No. 3-5 AI Act). Image or video models might be used to visualize safety aspects of high-risk products regulated under the New Legislative Framework (see Annex II Section A. AI Act). Unless any misuse can be verifiably technically excluded, LGAIMs will therefore generally count as high-risk systems.

This, however, entails that they have to abide by the high-risk obligations, in particular the establishment of a comprehensive risk management system, according to Article 9 AI Act. Setting up such a system seems to border on the impossible, given LGAIMs' versatility. It would compel LGAIM providers to identify and analyze all "known and foreseeable risks most likely to occur to health, safety and fundamental rights" concerning all possible high-risk uses of the LGAIM (Articles 9(2)(a), 4b(6) AI Act). On this basis, mitigation strategies for all these risks have to be developed and implemented (Article 9(2)(d) and (4) AI Act). Providers of LGAIMs such as ChatGPT would, therefore have to analyze the risks for every single, possible application in every single high-risk case contained in Annexes II and III concerning health, safety and all possible fundamental rights.

Similarly, performance, robustness, and cybersecurity tests will have to be conducted concerning all possible high-risk uses (Articles 15(1), 4b(6) AI Act). This seems not only almost prohibitively costly but also hardly feasible. The entire analysis would have to be based on an abstract, hypothetical investigation, and coupled with–again hypothetical–risk mitigation measures that will, in many cases, depend on the concrete deployment, which by definition has not been implemented at the moment of analysis. What is more, many of these possible use cases will, in the end, not even be realized because they are economically, politically, or strategically unviable. Hence, such a rule would likely create "much ado about nothing", in other words: a waste of resources. Ironically, the conception of Articles 4a-4c AI Act, as currently proposed, places a very high, and arguably undue, burden on providers of truly general-purpose AI systems. These providers will be most unlikely to be able to comply with the AI Act, by virtue of their model's sheer versatility–there will just be too many scenarios to contemplate. In conjunction with the proposed regime for AI liability, which facilitates claims for damages if the AI Act is breached, this also exposes LGAIM providers to significant liability risk [24, 77].

*3.1.3 Adverse Consequences for Competition*

Third, the current GPAIS rules would likely have significantly adverse consequences for the competitive environment surrounding LGAIMs. The AI Act definition specifically includes open source developers as LGAIM providers, of which there are several.[4] Some of these will explore LGAIMs not for commercial, but for philanthropic or research reasons. For example, Stable Diffusion was developed in a research project conducted at LMU Munich. While, according to its Article 2(7), the AI Act shall not apply to any (scientific, see Recital 12b AI Act) research and development activity regarding AI systems, this research exemption arguably does not apply anymore once the system is released into the wild, as any public release likely does not have scientific research and development as its "sole purpose" (Recital 12b AI Act), particularly when, as is often the case, a commercial partner enters to limit liability and provide necessary fine-tuning.

As a result, all entities–large or small–developing LGAIMs and placing them on the market will have to comply with the same stringent high-risk obligations. Given the difficulty to comply with them, it can be expected that only large, deep-pocketed players (such as Google, Meta, Microsoft/Open AI) may field the costs to release an approximately AI Act-compliant LGAIM. For open source developers and many SMEs, compliance will likely be prohibitively costly. Hence, the AI Act may have the unintended consequence of spurring further anti-competitive concentration in the LGAIM development market. This is in direct opposition to the spirit of Recital 61 Sentence 5 AI Act which–in the context of standardization–explicitly calls for an appropriate involvement of SMEs to promote innovation and competitiveness in the field of AI within the Union (see also Article 40(2)(b) and Article 53(1b)(a) AI Act). Similar effects have already been established concerning the GDPR [82]. In this sense, the AI Act threatens to undermine the efforts of the Digital Markets Act[5] to infuse workable competition into the core of the digital and platform economy.

*3.1.4 Critique of the European Parliament proposal*

In the European Parliament (EP), the question of how to regulate large generative AI models significantly delayed the formulation of the EP position on the AI Act. On February 7, 2023, it became known that the European Parliament intends to go even further than the Council by generally classifying generative AI systems as high-risk in Annex III of the AI Act [69], with the possible exception to use cases in which the output was subjected to proper human review. After a lengthy debate, a compromise was reached in late April/early May 2023, which did not include generative AI systems in Annex III.[6] Rather, the compromise foresees three layers of obligations that apply to generative AI systems [70, 83].

---

[4] See, e.g., https://www.kdnuggets.com/2022/09/john-snow-top-open-source-large-language-models.html.
[5] Regulation (EU) 2022/1925 of the European Parliament and of the Council of 14 September 2022 on contestable and fair markets in the digital sector, OJ L265/1 (DMA).

The first layer will apply to the providers (=developers) of a subset of GPAIS denominated "foundation models" (Article 28b(1)-(3) AI Act EP Version) and generative AI (Article 28b(4) AI Act EP Version). Referring to a well-known term in the computer science community [see, e.g., 40, 84], the EP version defines foundation models as an AI system "that is trained on broad data at scale, is designed for generality of output, and can be adapted to a wide range of distinctive tasks" (Article 3(1c) AI Act EP Version) [cf. also 40, at 3]. The focus on generality of output and tasks is indeed better suited to capture the specifics of large generative AI models than the vague definition of GPAIS (see Section 3.1.1). In line with suggestions made in this paper, the general obligations for all foundation models include data governance measures, particularly with a view to the mitigation of bias (Article 28b(2)(b) AI Act EP Version; see Section 4). Furthermore, appropriate levels of performance, interpretability, corrigibility, safety and cybersecurity must be maintained throughout the model's lifecycle. These requirements have to be tested for, documented, and verified by independent experts, Article 28b(2)(c) AI Act EP Version. Crucially, however, all foundation models also need to implement risk assessments, risk mitigation measures, and risk management strategies with a view to reasonably foreseeable risks to health, safety, fundamental rights, the environment, democracy and the rule of law, again with the involvement of independent experts, Article 28b(2)(a) AI Act EP Version. Effectively, this requirement is tantamount to classifying foundation models as high-risk per se.

A crucial element of the minimum standards for generative AI is contained in the "ChatGPT Rule" Art. 28b(4) AI Act EP Version. It contains three main elements. (i) The transparency obligation concerning the use of AI (Art. 28b(4) AI Act EP Version, Art. 52(1) AI Act) is a step in the right direction. It addresses obligations of providers towards users of AI systems. In our view, additionally, obligations of users towards recipients are warranted in some instances to fight the spread of fake news and misinformation (see Section 7.1). (ii) The rule on preventing a breach of EU law also arguably would benefit from greater detail. Here, the compliance mechanisms of the DSA should be transferred much more specifically, for example through clear, mandatory notice and action procedures and trusted flaggers (see Section 7.4). It goes without saying that the models must comply with applicable law. (iii) The disclosure of copyrighted material contained in training data may indeed help authors and creators enforce their rights. However, even experts often argue whether certain works are copyrightable at all or not. What must be avoided is that developers who have, e.g., processed 20 million images now have to conduct a full-scale legal due diligence on these 20 million images to decide for themselves whether they are copyrightable or not. Hence, it must therefore be sufficient to disclose, even in an over-inclusive manner, works which *may be* copyrightable, including those for which it is not clear whether they are ultimately copyrightable or not. Otherwise, again, practically prohibitive due diligence costs will arise. The individual author must then decide, when she discovers her work, whether she thinks it is protected by copyright or not.

The second layer refers to "new providers" which significantly modify the AI system, Art. 28(1)(b) and (ba) AI Act EP Version. This new provider, which is called deployer in our paper (see Section 3.2.2), assumes the obligations of the former provider upon substantial modification; the new provider takes on this role (Art. 28(1) and (2)(1) AI Act EP Version). The new rule on a fundamental rights impact assessment (Article 29a AI Act EP Version) also applies on this level of the concrete application. This rule also seems hardly operationalizable. Fundamental rights are a fuzzy category and difficult to implement at a technical level, where specific secondary regulation may be more useful (GDPR, non-discrimination law directives etc.). Importantly, it is also doctrinally misguided as private companies, in general, cannot violate fundamental rights of other private persons (fundamental rights bind the state, not the citizens; there are only some exceptions to that rule[7]). Also, what is the relationship to the general risk assessment (Articles 9 and 28b(1)(a))? A third layer of requirements relates to the AI value chain (Article 28(2)(2) AI Act EP Version), in line with suggestions made below in this paper (see Section 3.2.2).

---

[6] See note 2.

In our view, while containing steps in the right direction, this proposal would be ultimately unconvincing as it effectively treats foundation models as high-risk applications. Of course, as noted and discussed in detail below (Part 6), AI output may be misused for harmful speech and acts (as almost any technology). But not only does this seem to be rather the exception than the rule. The argument concerning adverse competitive consequences applies equally here. Under the EP version, risk assessment, mitigation, and management still remain focused on the model itself rather than the use-case specific application (Art. 28b(2)(a) and (f) AI Act EP Version), even though Recital 58a acknowledges that risks related from AI systems can stem from their specific use. Again, this leads to the onerous assessment and mitigation of hypothetical risks that may never materialize–instead of managing risks at the application level where the concrete deployment can be considered.

**3.2 Proposal: Focus on Deployers and Users**

This critique does not imply, of course, that LGAIMs should not be regulated at all. However, in our view, a different approach is warranted. Scholars have noted that the regulatory focus should shift [12, 13] and move towards LGAIM deployers and users, i.e., those calibrating LGAIMs for and using them in concrete high-risk applications. While some general rules, such as data governance, non-discrimination and cybersecurity provisions, should indeed apply to all foundation models (see Section 4), the bulk of the high-risk obligations of the AI Act should be triggered for specific use cases only and target primarily deployers and professional users.

*3.2.1 Terminology: Developers, Deployers, Users, and Recipients*

Lilian Edwards, for example, has rightly suggested to differentiate between developers of GPAIS, deployers, and end users [12, see also 24]. In the following, we take this beginning differentiation in the AI value chain one step further. In many scenarios, there will be at least four entities involved, in different roles [cf. 85]. We suggest that the terminology in the AI Act and other pertaining regulation must be adapted to the evolving AI value chain in the following way.

- **developer**: this is the entity originally creating and (pre-)training the model. In the AI Act, this entity is called the provider (under some further conditions, see Article 3(2)). Real-world examples would be OpenAI, Stability, or Google.
- **deployer**: this is the entity fine-tuning the model for a specific use case. The AI Act EP Version also uses the term, albeit in a slightly different manner, covering any person or entity using an AI system under its authority, except where the AI system is used in the course of a personal non-professional activity (Art. 3(4) AI Act EP Version); for the purposes of the AI Act EP Version, a deployer can be a (new) provider, Art. 28(2)(1) AI Act EP Version. Note that there could be several deployers (working jointly or consecutively), leading to a true AI value chain similar to OEM value chains. Alternatively, the developer could simultaneously act as a deployer (vertical integration) – just as for the purposes of the AI Act EP Version, a deployer can be a (new) provider, Art. 28(2)(1). This would raise the typical competition law issues of vertical integration [86-88].

---

[7] See, e.g., CJEU, Case C-414/16 (Egenberger) and following judgments in this line.

- **user**: this is the entity actually generating output from an LGAIM, e.g. via prompts, and putting it to use. The user may harness the output in a professional or a non-professional capacity. Hence, we introduce the following distinction within the category of users:
    - professional user: an entity using AI output for professional purposes, generally as defined in EU consumer law [89, 90]. The AI Act calls such entities professional users as well (cf. Article 2(8) AI Act). The professional user could be a for-profit or a non-profit company, an NGO, an administrative agency, a court, or the legislator, for example. Potential real-world examples would be the clothing and sportswear manufacturer from the first lead example, or any other entity from the groups of professional users just listed. Other professional users could be persons like the authors using ChatGPT or other generative AI systems for academic work and education. Note that, in addition, professional influencers should count as professional users when posting on social media; similarly, any individual making professionally motivated comments online would also count as a professional user in this respect. Finally, some exceptions from the EU consumer definition are in order: for example, employees[8] (and students, for that matter) should presumptively count as professional users when applying LGAIMs for job- or education-related tasks. Particularly concerning negative externalities of AI output, it should not matter whether users are pursuing a dependent or independent professional activity: any professional should adhere to compliance rules established in the AI Act to minimize third-party harm (e.g., Article 29 AI Act).
    - non-professional user: an entity using AI output for non-professional purposes, as defined in EU consumer law (with the proviso just discussed). The AI Act calls such entities non-professional users as well and exempts them from its (cf. Article 2(8) AI Act; the AI Act EP Version contains no general exemption, but excludes non-professionals from the definition of deployers, Article 3(4)). The parent from the lead example, who uses ChatGPT to brainstorm ideas for her daughter's birthday party, would fall into this category.

    If there was a deployer involved, the user would implement the fine-tuned model, calibrated by the deployer, into its product or use case. Typically, this would be a professional user. In the alternative, the user could harness the LGAIM by directly connecting with the developers, usually via an API. This includes non-professional users (e.g. generating birthday ideas via ChatGPT) as well as professional ones (e.g. writing an academic paper with the help of ChatGPT or drafting an advertisement campaign for a company with slogans designed by Luminous).

- **recipient**: this is the entity consuming the product offered by the user. It does not generate LGAIM output. Typically, it will be a consumer, but it could also be a company, an NGO, an administrative agency, a court, or the legislator. What characterizes the recipient in all these cases is that it sits at the receiving, passive end of the pipeline. In the terminology of the AI Act, these entities would arguably fall in the group of affected persons (see, e.g., Recitals 39 and 42 AI Act). Real-world examples include consumers exposed to AI-generated advertisements and children or parents invited with AI-designed birthday invitations.

With this terminology in place, regulatory obligations can be allocated to different types of actors in more nuanced ways. While developers should, to a certain extent, be subject to non-discrimination law and certain data governance provisions (Section 4), we suggest that the focus of regulatory duties should lie on deployers and users, for example concerning risk management systems (Art. 9 AI Act) or performance and robustness thresholds (Art. 15 AI Act) (see also below, Part 7).

---

[8] But see German Constitutional Court, Order of November 23, 2006, Case 1 BvR 1909/06: employees are consumers in the sense of the EU consumer law.

*3.2.2 The AI Value Chain*

Such a shift of the regulatory focus on deployers and users, however, entails several follow-up problems that need to be addressed [12]. First, deployers and users may be much smaller and less technologically sophisticated than LGAIM developers. This is not a sufficient reason to exempt them from regulation and liability, but it points to the importance of designing a feasible allocation of responsibilities along the AI value chain. Recent proposals discussed in the European Parliament point in this direction as well (see Section **Fehler! Verweisquelle konnte nicht gefunden werden.**). Obligations must be structured in such a way that deployers and users can reasonably be expected to comply with them, both by implementing the necessary technological adjustments and by absorbing the compliance costs.

Second, many of the AI Act's high-risk obligations refer to the training and modeling phase conducted, at least partially, by the LGAIM developers. Typically, LGAIM developers will pre-train a large model, which may then be fine-tuned by deployers, potentially in collaboration with developers [91, 92], while users ultimately make the decision what the AI system is used for specifically (e.g. commercial use for design or private use for generating an invitation text). To meet the AI Act requirements concerning training data (Article 10), documentation and record-keeping (Articles 11 and 12), transparency and human oversight (Articles 13 and 14), performance, robustness and cybersecurity (Article 15), and to establish the comprehensive risk management system (Article 9), any person responsible will need to have access to the developer's and deployer's data and expertise. This unveils a regulatory dilemma: focusing exclusively on developers entails potentially excessive and inefficient compliance obligations; focusing on deployers and users risks burdening those who cannot comply due to limited insight or resources.

Third, and related to the first and second aspect, individual actors in the AI value chain may simply not have the all-encompassing knowledge and control that would be required if they were the sole addressees of regulatory duties [93]. This more abstract observation also shows that shared and overlapping responsibilities may be needed.

In our view, the only way forward are collaborations between LGAIM providers, deployers and users with respect to the fulfillment of regulatory duties, where the regulator gives this (forced) collaboration adequate contours. More specifically, we suggest a combination of strategies known from pre-trial discovery, trade secrets law, and the GDPR. Under the current Council GA AI Act, such teamwork is encouraged in Article 4b(5): providers "shall" cooperate with and provide necessary information to users. A key issue, also mentioned in the Article, is access to information potentially protected as trade secrets or intellectual property (IP) rights [13, 94]. In this regard, Article 70(1) AI Act requires anyone "involved" in the application of the AI Act to "put appropriate technical and organizational measures in place to ensure the confidentiality of information and data obtained in carrying out their tasks and activities". To be workable, this obligation needs further concretization; the same holds true for the proposal by the European Parliament in this direction [95]; Art. 10(6a) AI Act EP Version only explicitly addresses a situation where such cooperation does not take place, and is limited to violations of Art. 10.

The problem of balancing collaboration and disclosure with the protection of information is not limited to the AI Act. In our view, it has an internal and external dimension. Internally, i.e., in the relationship between the party requesting and the party granting access, access rights are often countered, by the granting party, by reference to supposedly unsurmountable trade secrets or IP rights [96-98]. The liability directives proposed by the EU Commission, for example, contain elaborate evidence disclosure rules pitting the compensation interests of injured persons against the secrecy interests of AI developers and deployers [24, 77, 78]. Article 15(4) GDPR contains a similar provision, which by way of analogy also applies to the access right in Article 15(1) GDPR [99, 100].

Extensive literature and practical experience concerning this problem exists in the realm of the US pretrial discovery system [101-105]. Under this mechanism, partially adopted by the proposed EU evidence disclosure rules [24], injured persons may seek access to documents and information held by the potential defendant before even launching litigation. This, in turn, may lead to non-meritorious access requests by competitors. Such concerns are not negligible in the AI value chain. Here as well, developers, deployers and users may indeed not only be business partners but also be (potential) competitors. Hence, deployers' and users' access must be limited. Conversely, some flow of information must be rendered possible to operationalize compliance with high-risk obligations by deployers.

To guard against abuse, we suggest a range of measures. On the one hand, providers (and potentially deployers) may authorize the use of the model under the proviso that users sign *NDAs* and non-compete clauses. Private ordering should, to a certain extent, function between professional actors. On the other hand, it may be worthwhile to introduce provisions inspired by the US pretrial discovery system [98, 101, 106] and the proposed EU evidence disclosure mechanism (Article 3(4) AI Liability Directive, protective order). Hence, courts should be empowered to issue *protective orders*, which endow nondisclosure agreements with further weight and subject them to potential administrative penalties. The order may also exempt certain trade secrets from disclosure or allow access only under certain conditions (see F.R.C.P. Rule 26(c)(1)(G)). Furthermore, as the high-profile document review cases in the US concerning former and current US Presidents show, the appointment of a *special master* may, ultimately, strike a balance between information access and the undue appropriation of competitive advantage (cf. F.R.C.P. Rule 53(a)) [106]. With these safeguards in place, LGAIM developers should be compelled, and not merely encouraged, to cooperate with deployers and users concerning AI Act compliance if they have authorized the deployment.

Concerning the external dimension, the question arises of who should be responsible for fulfilling pertinent duties and be ultimately liable, regarding administrative fines and civil damages, if high-risk rules are violated. Here, we may draw inspiration from Article 26 GDPR (see also [12]). According to this provision, joint data controllers may internally agree on the bespoke allocation of GDPR duties (Article 26(1) GDPR), but remain jointly and severally liable (Article 26(3) GDPR). The reason for this rule is to facilitate data subjects' compensation, who must not fear to be turned away by both controllers with each blaming the other party. Moreover, the essence of the internal compliance allocation must be disclosed (Article 26(2) GDPR). This mechanism could, mutatis mutandis, be transferred to the AI value chain. Here again, collaboration is required and should be documented in writing to facilitate ex post accountability. Disclosing the core parts of the document, sparing trade secrets, should help potential plaintiffs choosing the right party for any ensuing disclosure of evidence requests under the AI liability regime. Finally, joint and several liability ensures collaboration and serves the compensation interests of injured persons. Internally, parties held liable by injured persons can then turn around and seek reimbursement from others in the AI value chain. For example, if the developers essentially retain control via an API distribution model, the internal liability burden will often fall on them. Developers' and deployers' liability, however, must end where their influence over the deployed model ends. Beyond this point, only the users should be the subject of regulation and civil liability (and vice versa, for example in control-via-API cases): incentives for action only make sense where the person incentivized is actually in a position to act [107, 108]. In the GDPR setting, this was effectively decided by the CJEU in the Fashion ID case (CJEU, C-40/17, para. 85). The sole responsibility of the users for certain areas should then also be included in the disclosed agreement to inform potential plaintiffs and foreclose non-meritorious claims against the developer and deployer. Such a system, in our view, would strike an adequate balance of interests and power between LGAIM developers, deployers, users, and affected persons.

Overall, the EP version of the AI Act now rightly contains rules on the AI value chain [83]. However, these need to be rendered more specific, as laid out in the preceding sections, to function effectively. Ultimately, allocating responsibility and liability along the value chain is crucial if the AI Act seeks to maintain its spirit of a technology-specific instrument that does not, however, regulate models per se, but primarily models in concrete use cases.

## 4 NON-DISCRIMINATION LAW

Some rules will have to apply directly to LGAIMs and LGAIM developers, however (see Section 7). A first candidate for such rules is non-discrimination law. Generally, it applies, in the US as well as the EU, in a technology-neutral way [4, 109-112, 113; but see also Massachusetts Legislature, Bill SD.1827 (193rd) for a (brief) proposal for anti-discrimination rules specifically for LGAIM operators]. Importantly, however, it only covers certain enumerated areas of activity, such as employment, education, or publicly available offers of goods and services [112, 114]. This begs the question whether general-purpose systems may be affected by non-discrimination provisions even before they have been deployed in specific use cases. Concerning EU law, the CJEU has held in a string of judgments (CJEU, Case C-54/07, Feryn; Case C-507/18, Associazione Avvocatura per i diritti LGBTI) that non-discrimination provisions may apply under certain conditions to preparatory activities preceding the primarily covered action, e.g., the actual job selection: in the concrete cases, statements made publicly and on a radio program indicating an intention not to recruit candidates of a particular sexual orientation, ethnic origin or race are considered "conditions for access to employment" if the relationship between the statement and the employer's recruitment policy is not merely hypothetical (CJEU, Case C-507/18, Associazione Avvocatura per i diritti LGBTI, para. 43; see also Case C-54/07, Feryn, para. 25). For anti-discrimination directives to apply, a preliminary measure must, therefore, concretely relate to an activity covered by the directives.

Regarding the (pre-)training of LGAIMs, the required link arguably exists if the model is specifically prepared for use (also) in discrimination-relevant scenarios (employment; education, publicly available goods or services etc.) [cf. also 25]. Conversely, if a generic LGAIM is developed without any specific link to such scenarios (even though it may be theoretically used in these cases) non-discrimination law generally does not apply to the development itself. However, it can become applicable if the development occurs with a view to offering the model to other actors (e.g., deployers, users) [112]. In this case, building the model may qualify as providing "goods and services, which are available to the public."[9] Furthermore, one could argue, in line with the Uber[10] and AirBnB[11] judgments of the CJEU, that LGAIM developers, in some cases, pre-determine the application of the model in concrete deployment cases [112, 115]. This would bar them from receding to a mere intermediary role.[12] Again, non-discrimination law obviously applies to the concrete deployment in the respective scenario (e.g., an employment use case). In all these cases, intricate questions ensue which transcend the scope of this paper, for example concerning concrete proof of discrimination, harm, and standing to sue. Ultimately, we would argue, however, that significant underperformance for legally protected groups will be indicative of (indirect or even direct) discrimination, establishing a prima facie case [4, 109-113].

Two final notes on the AI Act are in order here. First, sensible data governance rules should, in our view, apply to the foundation models themselves (see Section 7.3). Second, according to Article 29(4) AI Act, users must "implement human oversight and monitor the operation of the high-risk AI system[s]". In our view, this also implies screening for discriminatory biases in the output of the LGAIM. This obligation is severely limited in two respects, however. First, it only applies to high-risk AI systems; and, second, it only applies to professional users, Article 2(8) AI Act. The monitoring obligation should also extend to non-professional users: all users should be compelled to screen the output for evident cases of significant harm to third parties or collective interests, non-technically put, such as overtly discriminatory statements. In our view, such an obligation would not be overly burdensome for private users, if their duties are limited to

---

[9] Art. 3(1) of the Goods and Services Directive 2004/113/EC; see also Art. 3(1)(h) of the Race Equality Directive 2000/43/EC; Member State law contains protection on further grounds in some cases, e.g., in Germany (religion, disability, age or sexual orientation, § 2(1)(8) and § 19 AGG).
[10] CJEU, Case C-434/15, Asociación Profesional Elite Taxi.
[11] CJEU, Case C-390/18, AirBnB Ireland.
[12] Philipp is grateful to Sandra Wachter for pushing him on this point.

*best efforts* and to *evident* cases of *significant* harm to third parties. Notably, while varying by Member State, similar duties also apply when speaking for oneself, without aid by AI. The duty we propose may be more easily fulfilled sometimes (e.g., the prohibition of holocaust denial in Germany), more difficult in other instances (e.g., (non) gender-sensitive speech, currently hotly debated in the case law[13]). But at least, users should have to use best efforts to detect flagrant cases and, as a consequence, put developers on notice (Article 29(4)(2-3) AI Act), who are then called upon to act on the issue. With this compromise, lay users could be integrated into the enforcement of non-discrimination practice without overburdening them in standard usage of LGAIMs.

## 5  DATA PROTECTION UNDER THE GDPR AND THE ITALY BAN

A second major challenge for any AI model is GDPR compliance. While the requirements for large generative models are, arguably, not categorically different from those for any machine learning model, we refrain from an all-compassing substantiated analysis at this point.

However, we note that recent studies have shown that LGAIMs are vulnerable to inversion attacks, even to a greater extent than previous generative models, such as generative adversarial networks (GANs [116]) [28]. As a consequence, data used for training may be reproduced from the model. This is less of a concern for copyrighted material found in the training data by way of model inversion as the new text-and-data-mining exceptions (Articles 3 and 4 Copyright in the Digital Single Market Directive[14]), at least generally and where rightsholders have not opted out, allow the use of publicly accessible copyrighted material for machine learning purposes [117, 118]. However, a legal basis under Article 6 GDPR is necessary for any use of personal data for training [25, 27]. According to some scholars, even the model itself might be considered personal data, considering the possibility of inversion attacks [27]. As consent was typically not obtained [119], developers need to avail themselves of the balancing test and/or the purpose change test (Article 6(1)(f) and (4) GDPR) [120-123], potentially in conjunction with specific research exceptions (Article 89 GDPR) [124-126]. Compliance with the GDPR will depend on a range of factors [127-130], such as the intended purposes of the model, the type of personal data used, the likelihood of model inversion, and the probability of re-identification of concrete data subjects. While scholars are divided about whether the use of personal data for machine learning purposes should be permitted under the balancing test (permissive view: [25]; more restrictive view: [123, 131]; see also [132]), the threat of model inversion arguably weighs in favor of data subjects.

Importantly, if the personal data constitutes sensitive data in the sense of Article 9 GDPR, developers need to invoke, besides Article 6 GDPR, an exception under Article 9(2) GDPR. While there is no universal balancing test available at the EU level, some Member States have introduced (more restrictive) balancing tests for research purposes under the opening clauses of Article 9(2)(g-j) GDPR (see, e.g., § 22 of the German Data Protection Act, BDSG [25]). With respect to sensitive data, model inversion constitutes an even more serious threat to GDPR compliance as the risk of a reproduction of sensitive data will hardly be overcome by legitimate interests of developers, unless the model serves truly critical purposes (potentially in medicine or emergency cases).

Furthermore, chat interfaces of LGAIMs (such as ChatGPT) raise specific concerns, in particular regarding transparency. At the time of writing, i.e. May 2023, several enquiries are underway [133-138]. Most notably, the Italian

---

[13] See, e.g., Administrative Court of Berlin, Order of March 24,2023, Case 3 L 24/23 (gender sensitive language at schools allowed if not taught as mandatory); German Federal Private Law Court [BGH], Judgment of March 13, 2018, Case VI ZR 143/17 (usage of generically masculine grammatical form in German not an infringement of general personality rights of women or of German/EU non-discrimination law); see also German Constitutional Court, Order of May 26, 2020, Case 1 BvR 1074/18.
[14] Directive (EU) 2019/790 of the European Parliament and of the Council of 17 April 2019 on copyright and related rights in the Digital Single Market, OJ L 130, 17.5.2019, 92.

Data Protection Authority (Garante per la Protezione dei Dati Personali - GPDP) even imposed a temporary limitation on the processing of Italian users' data by OpenAI [139]. For a few weeks, OpenAI geoblocked ChatGPT in Italy to conform with the order [140]. To us, this order seems a timely call for more transparency: OpenAI was–and arguably still is, to a lesser extent–secretive about what training data it has used,[15] and about how input by users of ChatGPT is processed.[16] This issue may (also) be tackled within the framework of the GDPR, as different provisions apply and demand transparency.

For data that was collected from the internet and used in training, Article 14 GDPR may apply. This provision governs transparency where personal data have not been obtained from the data subject. However, in many cases, providing the data subject with the information as required by Article 14(1) and (2) GDPR to individual persons whose data were part of training data may require a disproportionate effort–and thus may not be required, Article 14(5)(b) GDPR [141, para. 62 et seqq.]. According to Recital 62 GDPR, "the number of data subjects, the age of the data and any appropriate safeguards adopted should be taken into consideration". As the Article 29 Working Party highlights, disproportionate effort arises particularly if the data was collected from a large number of individuals whose contact details are unknown [141, para. 63, example]. However, the model construction, due to its commercial nature, may not fall under the research and statistics clause of Article 89 GDPR (which would speak in favor of disproportionate effort, Rec. 62 and 159). On the other hand, data subjects have a significant interest in being informed. Arguably, training data scraping and processing constitutes a borderline case under Article 14(5) GDPR. In our view, the controller (e.g., OpenAI, see Article 4(7) GDPR) should at least have to document the balancing exercise conducted under Article 14(5) GDPR. This obligation arguably is an essential part of the accountability required by Article 5(2) GDPR [141, para. 64]. Moreover, the controller should disclose the way in which it collected training data in a publicly available document.

By contrast, no exception comparable to Article 14(5) GDPR applies for the processing of personal data that users provide within a chat interface. Here, Article 13 GDPR governs and requires information to be provided to the data subject, in particular: the purposes of the processing for which the personal data are intended as well as the legal basis for the processing and, in the context of Article 6(1)(f) GDPR, the legitimate interests pursued in data processing. So far, no such information has been provided to users by OpenAI, for example.

Moreover, the order expresses concern with the "absence of a suitable legal basis" for the processing of data (particularly regarding minors, for whose consent Article 8 GDPR establishes specific rules). This concern can be expected to be subject to further debate in view of Article 6(1)(f) GDPR, see above. Last, the GPDP also emphasizes that the propensity of LGAIMs to hallucinate may also raise GDPR issues. Article 5(1)(d) GPDR requires personal data to be accurate. LGAIM output that entails incorrect personal data would therefore also contravene the principle of accuracy. Beyond that, discriminatory output will violate Art. 5(1)(a) GDPR, the fairness principle [112].

After negotiations with OpenAI, the GPDP issued another order, according to which the provisional suspension would be lifted if OpenAI complied with certain requirements[142]. By the end of April 2023, OpenAI was allowed to reinstate its services [143]. In the meantime, it had implemented certain measures [143]: most prominently, OpenAI provides information on its website describing the way it processes data and on corresponding rights of users and non-users and has clarified that for the processing of users' data will be consent (Article 6(1)(a) GDPR) or legitimate interest (Article 6(1)(f) GDPR).[17] Furthermore, users can exercise their rights (to access, delete or correct personal information) or object to the processing of their data via their ChatGPT account or via a publicly accessible form. As regards the protection of minors, OpenAI has established an age gate and age verification tools. Regarding specific points, the procedure raises pressing questions: Was the ban an adequate measure–even though Article 58(1) GDPR provides for less drastic alternatives and even though inquiries were still ongoing? And if so, were the measures sufficient to alleviate the GDPR concerns raised beforehand? In particular, is the age gate effective and is the information provided sufficient to comply with the information duties of Articles 13 and 14 GDPR? Also, it is not even clear whether the form provided for objections as a HubSpot form (as of May 5, 2023) complies with the GDPR.

---

[15] See, e.g. OpenAI, GPT-4 Technical Report, https://cdn.openai.com/papers/gpt-4.pdf (accessed on April 4, 2023).
[16] See OpenAI, Terms of use (March 13, 2023), https://openai.com/policies/terms-of-use (accessed on April 4, 2023), under 3.
[17]  https://openai.com/policies/privacy-policy and https://help.openai.com/en/articles/7842364-how-chatgpt-and-our-language-models-are-developed (accessed on May 5, 2023).

Still, overall, the decisions by the Italian GPDP rightly point to the legitimate interests, and rights, of data subjects to be informed about how their personal data is used in training and fine-tuning generative AI models. It should be taken as a welcome wake-up call to the community of developers to share crucial information–on training, personal data, and pertinent risks–with the general public, instead of guarding secrets under the misnomer of OpenAI et al.

## 6 GENERATIVE MODEL CONTENT MODERATION: THE EUROPEAN DIGITAL SERVICES ACT

The third large regulatory frontier concerning LGAIMs is content moderation. Generative models, as virtually any novel technology, may be used for better (think: birthday cards) or worse purposes (think: shitstorm) [144]. The developers of ChatGPT, specifically, anticipated the potential for abuse and trained an internal AI moderator, with controversial help from Kenyan contractors [67] , to detect and block harmful content [12]. AI Research has made progress in this area recently [145-147]. OpenAI has released a content filtering mechanism which users may apply to analyze and flag potentially problematic content along several categories (violence; hate; sexual content etc.).[18] Other large generative models have similar functionalities. However, actors intent on using ChatGPT, and other models, to generate fake or harmful content will find ways to prompt them to do just that. Prompt engineering is becoming a new art to elicit any content from LGAIMs [148] and fake news is harder to detect than hate speech, even though industry efforts are underway via increased model and source transparency [149]. As could be expected, DIY instructions for circumventing content filters are already populating YouTube and reddit,[19] and researchers have already generated an entire hate-filled shitstorm, along with code for proliferation, using ChatGPT [8]. The propensity of ChatGPT particularly to hallucinate when it does not find ready-made answers can be exploited to generate text devoid of any connection to reality, but written in the style of utter confidence, persuasion, scholarly attitude, or derision, as conditions warrant. While the European Parliament is investigating LENSA for inappropriate content generation [150, 151] the timing of the advent of truly powerful LGAIMs could hardly be any more favorable for malicious actors. The unprovoked Russian attack on Ukraine, the Corona pandemic, climate change, and the political feuds in the US and beyond already fuel hate crimes and fake news [152, 153]. This toxic political climate now meets the factual demolition of content moderation on Twitter under the auspices of its new owner. LGAIMs could well be a powerful instrument in the upcoming election cycles to target individual actors and sway public opinion. They allow for the automated mass production, and proliferation, of highly sophisticated, seemingly fact-based, but actually utterly nonsensical fake news and harmful speech campaigns.

To stem the tide of such phenomena, the EU has recently enacted the DSA. However, LGAIMs were not in the focus of public attention at the time when the DSA was being drafted. Hence, the DSA was designed to mitigate illegal content on social networks, built by human actors or the occasional Twitter bot, not to counter LGAIMs. The problem lies not in the territorial applicability of the provision: the DSA, like the AI Act, covers services offered to users in the EU, irrespective of where the providers have their place of establishment (Article 2(1), 3 (d) and (e) DSA). Platforms like Facebook or Twitter must implement a notice and action system where users can report potentially illegal content that will then be reviewed and removed if found illegal (Article 16 DSA). Larger platforms must have an internal complaint and redress system (Article 20 DSA) and provide out-of-court dispute resolution (Article 21 DSA). Hence, users are spared the lengthy process of going to court to challenge problematic content. Repeat offenders of content moderation policies risk having their accounts suspended (Article 23 DSA). Content highlighted as problematic by so-called trusted flaggers, which have to register with Member States, must be prioritized and decided upon without undue delay (Article 22 DSA). Very large online platforms (> 45 million active users) must also implement a comprehensive compliance system, including proactive risk management and independent audits (Articles 33-37 DSA).

---

[18] See https://platform.openai.com/docs/api-reference/moderations.
[19] See, e.g., https://www.youtube.com/watch?v=qpKlnYLtPjc; https://www.reddit.com/r/OpenAI/comments/zjyrvw/a_tutorial_on_how_to_use_chatgpt_to_make_any/.

Passed with the best of intentions, the DSA, however, seems outdated at the moment of its enactment. This results from two crucial limitations in its scope of application. First, it covers only so-called intermediary services (Article 2(1) and (2) DSA). Article 3(g) DSA defines them as "mere conduit" (e.g., Internet access providers), "caching" or "hosting" services (e.g., social media platforms, see also Recital 28 DSA). Arguably, however, LGAIMs do not fall into any of these categories. Clearly, they are not comparable to access or caching service providers, which power Internet connections. Hosting services, in turn, are defined as providers storing information provided by, and at the request of, a user (Article 3(g)(iii) DSA) [see also 154]. While users do request information from LGAIMs via prompts, they can hardly be said to provide this information. Rather, other than in traditional social media constellations, it is the LGAIM, not the user, who produces the text. To the contrary, CJEU jurisprudence shows that even platforms merely storing user-generated content may easily lose their status as hosting providers, and concomitant liability privileges under the DSA (and its predecessor in this respect, the E-Commerce Directive), if they "provide assistance" and thus leave their "neutral position", which may even mean merely promoting user-generated content (CJEU, Case C-324/09, L'Oréal para 116). A fortiori, systems generating the content themselves cannot reasonably be qualified as hosting service providers. Hence, the DSA does not apply.

This does not imply that LGAIM content generation is not covered by content liability laws. Rather, its output may be covered by speech regulation, similar to comments made by human users online. However, this branch of the law is largely left to Member State tort law, with the exception of Article 82 GDPR in the case of processing personal data of victims, which seems rather far-fetched in LGAIM constellations. Not only does such direct speech regulation vary considerably between Member States [155]: one Member State (Germany) even declined to enforce a judgment of another Member States' court (Poland) on speech regulation, as it saw a collision with its constitutional provisions on free speech, amounting to a breach of its own *ordre public*[20]. It also often lacks precisely the instruments the DSA has introduced to facilitate the rapid yet procedurally adequate removal of harmful speech and fake news from the online world: notice and action mechanisms flanked by procedural safeguards; trusted flaggers; obligatory dispute resolution; and comprehensive compliance and risk management regimes for large platforms.

The risk of a regulatory loophole might be partially closed, one might object, by the applicability of the DSA to LGAIM-generated posts that human users, or bots, publish on social networks. Here, the DSA generally applies, as Twitter et al. qualify as hosting service providers. However, a second important gap looms: Recital 14 DSA specifies that the main part of the regulation does not cover "private messaging services." While the notice and action mechanism applies to all hosting services, instruments like trusted flaggers, obligatory dispute resolution, and risk management systems are reserved for the narrower group of "online platforms" [156]. To qualify, these entities must disseminate information to the public (Article 3(g)(iii), (k) DSA). According to Recital 14 DSA, closed groups on WhatsApp and Telegram, on which problematic content particularly proliferates, are explicitly excluded from the DSA's online platform regulation (Articles 19 et seqq. DSA) as messages are not distributed to the general public. With the right lines of codes, potentially supplied by an LGAIM as well [8], malicious actors posting content in such groups may therefore fully escape the ambit, and the enforcement tools, of the DSA.

---

[20] German Federal Private Law Court [BGH],, Judgment of 19 July 2018, Case IX ZB 10/18.

Hence, the only action to which the full range of the DSA mechanisms continues to apply is the posting of LGAIM-generated content on traditional social networks, such as Twitter, YouTube, or Instagram. However, at this point in time, Pandora's box has already been opened. Misinformation may also be spread effectively and widely via interpersonal communication. Even if the EU legislator has decided to exclude closed groups from the scope of the DSA [157], this balance needs to be reassessed in the context of readily available LGAIM output, which exacerbates risks. Even the most stringent application of DSA enforcement mechanisms, potentially coupled with GDPR provisions on erasure of data (Article 17(2) and 19 GDPR), cannot undo the harm done, and often cannot prevent the forward replication of problematic content [158]. Overall, current EU law, despite the laudable efforts in the DSA to mitigate the proliferation of fake news and hate speech, fails to adequately address the dark side of LGAIMs.

## 7 POLICY PROPOSALS

The preceding discussion has shown that regulation of LGAIMs is necessary, but must be better tailored to the concrete risks they entail. Hence, we suggest a shift away from the wholesale AI Act regulation envisioned in the general approach of the Council of EU toward specific regulatory duties and content moderation. Importantly, regulatory compliance must be feasible for LGAIM developers large and small to avoid a winner-takes-all scenario and further market concentration [82]. This is crucial not only for innovation and or consumer welfare [33, 159, 160], but also for environmental sustainability. While the carbon footprint of IT and AI is significant and steadily rising [54-58], and training of LGAIMs is particularly resource intensive [161], large models may ultimately create fewer greenhouse gas emissions than their smaller brethren if they can be adapted to multiple uses.

Against this background, we envision three layers of obligations for LGAIMs: the first set of minimum standards for all LGAIMs; a second set of specific high-risk rules applying only to LGAIMs used in concrete high-risk use cases; and the third set of rules governing collaboration along the AI value chain (see Section 3.2.2) to enable effective compliance with the first two sets of rules.

Concerning minimum standards, first and foremost, the EU acquis applies to developers of LGAIMs as well, putting the GDPR (see Section 5), non-discrimination law (Section 4), as well as product liability [24] center stage. In addition, transparency rules, now also proposed by the European Parliament [70], must apply (see below, Section 7.1). Furthermore, specific risks of such outstanding relevance that they should be addressed at the upstream level, rather than delegated to deployers in specific use cases, must be allocated to developers as part of the minimum standards. This concerns, in our view, selected data governance duties (Art. 10 AI Act, see Section 4) and rules on the ever more important issue of cybersecurity (Art. 15 AI Act). Finally, sustainability rules [24] as well as content moderation (see below, Section 7.4) should also form part of the minimum standards applicable to all LGAIMs.

In the following, we make four concrete, workable suggestions for LGAIM regulation on the first and second level: (i) transparency obligations (first and second level); (ii) mandatory yet limited risk management (second level); (iii) non-discrimination data audits; and (iv) expanded content moderation.

## 7.1 Transparency

The AI Act contains a wide range of disclosure obligations (Article 11, Annex IV AI Act) that apply, however, only to high-risk systems. In our view, given the vast potential and growing relevance of LGAIMs for many sectors of society, LGAIMs should — irrespective of their categorization as high-risk or non-high-risk — be subject to two distinct transparency duties.

*7.1.1 Transparency requirements for developers and deployers*

First, *LGAIM developers and deployers* should be required to report on the provenance and curation of the training data, the model's performance metrics, and any incidents and mitigation strategies concerning harmful content. Ideally, to the extent technically feasible [54, p. 28, Annex A], they should also disclose the model's greenhouse gas (GHG) emissions, to allow for comparison and analysis by regulatory agencies, watchdog organizations, and other interested parties. This information could also serve as the basis for an AI Sustainability Impact Assessment [24, p. 65 f., see also 162].

*7.1.2 Transparency requirements for users*

Second, *professional users* should be obligated to disclose which parts of their publicly available content were generated by LGAIMs, or adapted based on their output. Specifically, this entails that in adidas example, adidas needs to adequately inform users that the design was generated using, e.g., Stable Diffusion. While the added value of such information may be limited in sales cases, such information is arguably crucial in any cases involving content in the realm of journalism, academic research, or education. Here, the recipients will benefit from insight into generation pipeline. They may use such a disclosure as a warning signal and engage in additional fact checking or to at least take the content *cum grano salis*. Eventually, we imagine differentiating between specific use cases in which AI output transparency vis-à-vis recipients is warranted (e.g., journalism, academic research or education) and others where, based on further analysis and market scrutiny, such disclosures may not be warranted (certain sales, production and B2B scenarios, for example). For the time being, however, we would advocate a general disclosure obligation for professional users to generate further information and insight into the reception of such disclosures by other market participants or recipients.

Conversely, we submit that *non-professional users* should not be required to inform about the use of AI. In the birthday example, hence, a parent would not need to inform the parents that the invitation or the entire design of the birthday party was rendered possible by, e.g., Aleph Alpha's Luminous or ChatGPT. One might push back against this in cases involving the private use of social media, particularly harmful content generated with the help of LGAIMs. However, any rule to disclose AI-generated content would likely be disregarded by malicious actors seeking to post harmful content. Eventually, however, one might consider including social media scenarios into the domain of application of the transparency rule if AI detection tools are sufficiently reliable. In these cases, malicious posts could be uncovered, and actors would face not only the traditional civil and criminal charges, but additionally AI Act enforcement, which could be financially significant (administrative fines) and hence create even greater incentives to comply with the transparency rule, or refrain from harmful content propagation.

The enforcement of any user-focused transparency rule being arduous, it must be supported by technical measures such as digital rights management and watermarks imprinted by the model [163]. The European Parliament is currently pondering a watermark obligation for generative AI [162]. Importantly, more interdisciplinary research is necessary to develop markings that are easy to use and recognize, but hard to remove by average users [164]. This should be coupled with research on AI-content detection to highlight such output where watermarks fail [146, 165].[21]

---

[21] See also https://openai.com/blog/new-ai-classifier-for-indicating-ai-written-text/.

## 7.2 Risk Management and Staged Release

As mentioned, one major obstacle to the effective application of the AI Act to LGAIMs proper is comprehensive risk management. Here, novel approaches are needed. Scholars have rightly suggested that powerful models should be released consciously, trading off the added benefit of public scrutiny with the added risk of misuse in the case of full public releases [84, 166]. Additional factors, such as the balance of power among developers, must also be considered [166]. In our view, a limited, staged release, coupled with only access for security researchers and selected stakeholders, may often be preferable [see also 9, 84, 167, 168]. This adds a nuanced, community-based risk management strategy by way of codes of conduct to the regulatory mix [cf. also 168]. Regulatory oversight could be added by way of "regulated self-regulation;" an approach with potentially binding effect of the code of conduct, à la Article 40 GDPR, seems preferable to the purely voluntary strategy envisioned in Article 69 AI Act.

Importantly, the full extent of the high-risk section of the AI Act, including formal risk management, should only apply if and when a particular LGAIM (or GPAIS) is indeed used for high-risk purposes (see Part 0). This strategy aligns with a general principle of product safety law [13]: not every screw and bolt must be manufactured to the highest standards. For example, only if they are used for spaceships, stringent product safety regulations for producing aeronautics material apply[22]–but not if they are sold in the local DIY store for generic use. The same principle should be applied to LGAIMs.

## 7.3 Non-Discrimination and training data

Furthermore, we suggest that, as an exception to the focus on LGAIM deployers, certain data curation duties, for example representativeness and approximate balance between protected groups (cf. Article 10 AI Act), should apply to LGAIM developers. Discrimination, arguably, is too important a risk to be delegated to the user stage and must be tackled during development and deployment. Wherever possible, discrimination AI systems should be addressed at its roots (often the training data) and not propagated down the ML pipeline or AI value chain. After all, discriminatory output should, in our view, be avoided in all use cases, even on birthday cards. The regulatory burden, however, must be adapted to the abstract risk level and the compliance capacities (i.e., typically the size) of the company. For example, LGAIM developers should have to pro-actively audit the training data set for misrepresentations of protected groups, in ways proportionate to their size and the type of training material (curated data vs. Twitter feeds scraped from the Internet), and implement feasible mitigation measures. At the very least, real-world training data ought to be complemented with synthetic data to balance historical and societal biases contained in online sources. For example, content concerning professions historically reserved for one gender (nurse; doctor) could be automatically copied and any female first names or images exchanged by male ones, and vice versa, creating a training corpus with more gender-neutral professions for text and image generation.

## 7.4 Content Moderation

One of the biggest challenges for LGAIMs is, arguably, their potential misuse for disinformation, manipulation, and harmful speech. In our view, the DSA rules conceived for traditional social networks must be expanded and adapted accordingly.

### 7.4.1 Selective expansion of the DSA to LGAIMs

The European Parliament has partially addressed this challenge by stipulating that foundation models must not violate EU law [76]. In our view, however, regulation should go one step further by selectively expanding DSA rules to LGAIM developers and deployers. LGAIMs, and society, would benefit from mandatory notice and action mechanisms, trusted flaggers, and comprehensive audits for models with particularly many users. The regulatory loophole is particularly virulent for LGAIMs offered as standalone software, as is currently the case. In the future, one may expect an increasing integration into platforms of various kinds, such as search engines or social networks, as evidenced by LGAIM development or acquisition by Microsoft, Meta, or Google. While the DSA would then technically apply, it would still have to be updated to ensure that LGAIM-generated content is covered just like user-generated content. In particular, as LGAIM output currently is particularly susceptible to being used for the spread of misinformation, it seems advisable to require LGAIM-generated content to be flagged as such–if technically feasible. Doctrinally, this could be achieved via an amendment of the DSA or of Article 29 AI Act, which already contains notification duties in its para. 4 (see Part. 4). Given the current political process in the EU, the latter option seems more realistic.

---
[22] See, e.g., product standards, aerospace series, DIN EN 4845–4851 (December 2022) on screws.

*7.4.2 Implementation in practice*

How could DSA-style content moderation applied to ChatGPT et al. look like in practice? We envision it to have two components. These components would combine centralized and decentralized monitoring within a notice-and-action mechanism (cf. Article 16 DSA).

The first component harnesses the wisdom of the crowd, as it were, to correct LGAIM output. Users should be enabled to flag problematic content and give notice. A special status should be given to a specific group of users, trusted flaggers (cf. Article 22 DSA), who could be private individuals, technologies savvy NGOs, or volunteer coders. After registering with the competent authority, they would essentially function as a decentralized content monitoring team. They could experiment with different prompts and see if they manage to generate harmful or otherwise problematic content. They could also scan the internet for tools to circumvent content moderation policies and instruments at LGAIMs.[23] If they find something, trusted flaggers would send a notice containing the prompt and the output to a content moderation check-in point of the respective LGAIM system, which would forward the notice to developers and/or deployers.

Here, the second component enters the scene, geared toward tech engineers working with developers or deployers. They would have to respond to notices; those submitted by trusted flaggers would have to be prioritized by the content moderation team. Their job, essentially, is to modify the AI system, or to block its output, so that the flagged prompt does not generate problematic output anymore, and to generally search for ways to block easy workarounds likely tried by malicious actors. Furthermore, if the LGAIM system is large enough, they would be tasked with establishing a more comprehensive compliance system (cf. Article 34-35 DSA). Overall, such a combination of centralized and decentralized monitoring could prove more effective and efficient than current systems relying essentially on goodwill to handle the expected flood of hate speech, fake news and other problematic content generated by LGAIMs.

*7.4.3 Avenues for future fundamental research*

Finally, like in the other areas, more fundamental research is additionally needed to not only mark AI-generated content, but to integrate adherence to facts and content moderation into the models themselves. Transparent LGAIMs connected to (good-old) knowledge bases, such as AtMan [43], present a promising way forward in this direction. Here again, technology will not "fix" the problem on its own; rather, technical advances must be embedded into societal discourse, and sensible, not necessarily technology-specific regulation, to reap the benefits of LGAIMs while mitigating its risks.

---

[23] See Fn. 19.

## 7.5 Outlook: Technology-specific vs. technology-neutral regulation

Overall, we have added several policy proposals. As a matter of regulatory technique, the legislator should, in our view, strive to shift its strategy from technology-specific regulation–which will often be outdated before eventually enacted– toward more technology-neutral regulation wherever possible. Due to space constraints, we cannot elaborate on this point. However, the previous discussion has shown that non-discrimination law, formulated in a technology-neutral way, continues to grapple with various challenges, but arguably does a better job capturing the dynamics of LGAIM development than the AI Act or the DSA, at least in the way they are currently enacted and proposed. While technology-neutral regulation must be tailored, via agency decisions, regulatory guidelines, and court judgments, to specific technologies, such "small-scale" adaptations are, arguably, often faster to produce than changes to a formal, technology-specific, legislative act. For example, to extend the DSA to LGAIMs in specific ways, one would have to update the DSA or include a reference in the AI Act. Both modifications require concurring decisions by the European Parliament and the Council (Article 289 TFEU). In non-discrimination law, by contrast, all that is needed, in principle, is an adequate interpretation of existing law by agencies and courts. Their decisions, at least in lower courts, can potentially be rendered faster and be used more flexibly to carve out (preliminary[24]) safe harbors for developers, deployers, and users, and to establish red lines to protect affected persons.

## 8 CONCLUSION

Scholars and regulators have long suggested that technology-neutral laws may be better prepared to tackle emerging risks given the rapid pace of innovation in machine learning [169-171]. While this claim, arguably, cannot be generally affirmed or refuted, LGAIMs offer a cautionary example for regulation focused specifically on certain technologies. As our study shows, technology-neutral laws sometimes fare better because technology-specific regulation (on platforms; AI systems) may be outdated before (AI Act, AI liability regime) or at the moment of its enactment (DSA). Overall, we add several policy proposals to the emerging regulatory landscape surrounding LGAIMs.

To start with, we argue for a new, differentiated terminology to capture the relevant actors in the AI value chain, in LGAIM settings and beyond. These include: LGAIM developers, deployers, professional and non-professional users, as well as recipients of LGAIM output. Such a nuanced understanding is necessary to allocate regulatory duties to specific actors and activities in the AI value chain. The general approach adopted by the Council of the EU failed to address the specificities of the LGAIM value chain. Rules in the AI Act and other direct regulation must match the specificities of pre-trained models.

More concretely, we propose three layers of rules applicable to LGAIMs. The first layer applies directly to all LGAIMs. It comprises existing, technology-neutral regulation such as the GDPR or non-discrimination provisions. Arguably, a version of Art. 10 AI Act and of the cybersecurity rules in Art. 15 AI Act should also apply to LGAIM developers. Furthermore, sustainability and content moderation instruments also form part of this first layer. Art. 28b AI Act EP Version represents an imperfect step into this direction.

On the second layer, we suggest generally singling out concrete high-risk applications, and not the pre-trained model itself, as the object of high-risk obligations. For example, it seems inefficient and practically infeasible to compel the developers of ChatGPT et al. to draw up a comprehensive risk management system covering, and mitigating, all the hypothetical risks to health, safety and fundamental rights such LGAIMs may pose – as the AI Act EP Version still does (Art. 28b(1)(a) and (f)). Rather, if used for a concrete high-risk purpose (e.g., summarizing or grading résumés in employment decisions), the specific deployer and user should have to comply with the AI Act's high-risk obligations, including the risk management system.

---

[24] Ultimately, we agree that it may take a substantial amount of time for final decisions to emerge from the court system. Only these can deliver a higher degree of legal certainty. However, even lower-court judgments or agency decisions may, arguably, indicate useful directions and, at least, be used to model compliance tools accordingly, even if policies may have to be revised if decisions are reversed in higher instances.

The devil, however, is in the detail: providers need to cooperate with deployers to comply with even such narrower regulatory requirements. Hence, a third layer mandating collaboration between actors in the AI value chain for compliance purposes is necessary. Here, we suggest drawing on experience from the US pretrial discovery system and Art. 26 GDPR to balance interests in the access to information with trade secret protection. Art. 28(2a) AI Act EP Version has partly taken up this proposal.

The last section makes concrete policy proposals. For example, detailed transparency obligations are warranted. This concerns both LGAIM developers and deployers (performance metrics; harmful speech issues arisen during pre-training) as well as users (disclosure of the use of LGAIM-generated content).

Finally, the core of the DSA's content moderation rules should be expanded to cover LGAIMs. Art. 28b(4)(b) and generative AI (Article 28b(4) AI Act EP Version) moves in this direction. More specifically, however, rules must also include notice and action mechanisms, trusted flaggers, and, for very large LGAIM developers, comprehensive risk management systems and audits concerning content regulation. Arguably, it is insufficient to tackle AI-generated hate speech and fake news ex post, once they are posted to social media. At this point, their effect will be difficult to stop. Rather, AI generation itself must be moderated by an adequate combination of AI tools, developer and user interventions, and law.

In all areas, regulators and lawmakers need to act fast to keep track with the unchained dynamics of GPT-4 et al. Updating regulation is necessary both to maintain the civility of online discourses and to create a level playing field for developing and deploying the next generation of AI models, in the EU and beyond.

## ACKNOWLEDGMENTS

Passages taken over from ChatGPT are found in the section on technical foundations, all italicized and marked with "", and referenced by the prompt used. They were all collected on January 17, 2023. We deem them factually correct unless otherwise noted. This paper benefitted from comments by Johannes Otterbach, Sandra Wachter, and audiences at AI Campus Berlin, Bucerius Law School (Hamburg), Magdalen College (Oxford), University of Hamburg, and Weizenbaum Institute of the Connected Society. All errors remain entirely our own.

## 9 REFERENCES

[1] Glaese, A., McAleese, N., Trębacz, M., Aslanides, J., Firoiu, V., Ewalds, T., Rauh, M., Weidinger, L., Chadwick, M. and Thacker, P. Improving alignment of dialogue agents via targeted human judgements. *arXiv preprint arXiv:2209.14375* (2022).
[2] Shuster, K., Xu, J., Komeili, M., Ju, D., Smith, E. M., Roller, S., Ung, M., Chen, M., Arora, K. and Lane, J. Blenderbot 3: a deployed conversational agent that continually learns to responsibly engage. *arXiv preprint arXiv:2208.03188* (2022).
[3] Scao, T. L., Fan, A., Akiki, C., Pavlick, E., Ilić, S., Hesslow, D., Castagné, R., Luccioni, A. S., Yvon, F. and Gallé, M. Bloom: A 176b-parameter open-access multilingual language model. *arXiv preprint arXiv:2211.05100* (2022).
[4] Zuiderveen Borgesius, F. J. Strengthening legal protection against discrimination by algorithms and artificial intelligence. *The International Journal of Human Rights*, 24, 10 (2020), 1572-1593.

[5] Lee, D. and Yoon, S. N. Application of artificial intelligence-based technologies in the healthcare industry: Opportunities and challenges. *International Journal of Environmental Research and Public Health*, 18, 1 (2021), 271.
[6] Aung, Y. Y., Wong, D. C. and Ting, D. S. The promise of artificial intelligence: a review of the opportunities and challenges of artificial intelligence in healthcare. *British medical bulletin*, 139, 1 (2021), 4-15.
[7] Marcus, G. *A Skeptical Take on the A.I. Revolution*. The Ezra Klein Show, The New York Times, City, 2023.
[8] Beuth, P. *Wie sich ChatGPT mit Worten hacken lässt*. Der Spiegel, City, 2023.
[9] Bergman, A. S., Abercrombie, G., Spruit, S., Hovy, D., Dinan, E., Boureau, Y.-L. and Rieser, V. *Guiding the release of safer E2E conversational AI through value sensitive design*. Association for Computational Linguistics, City, 2022.
[10] Mirsky, Y., Demontis, A., Kotak, J., Shankar, R., Gelei, D., Yang, L., Zhang, X., Pintor, M., Lee, W. and Elovici, Y. The threat of offensive ai to organizations. *Computers & Security* (2022), 103006.
[11] Satariano, A. and Mozur, P. *The People Onscreen Are Fake. The Disinformation Is Real.*, City, 2023.
[12] Edwards, L. Regulating AI in Europe: four problems and four solutions (2022), 2022.
[13] Hacker, P., Engel, A. and List, T. *Understanding and regulating ChatGPT, and other large generative AI models*. City, 2023.
[14] Gutierrez, C. I., Aguirre, A., Uuk, R., Boine, C. C. and Franklin, M. A Proposal for a Definition of General Purpose Artificial Intelligence Systems. *Working Paper, https://ssrn.com/abstract=4238951* (2022).
[15] Heikkilä, M. *The EU wants to regulate your favorite AI tools*. City, 2023.
[16] KI-Bundesverband *Large European AI Models (LEAM) as Leuchtturmprojekt für Europa*. City, 2023.
[17] Goldstein, J. A., Sastry, G., Musser, M., DiResta, R., Gentzel, M. and Sedova, K. Generative Language Models and Automated Influence Operations: Emerging Threats and Potential Mitigations. *arXiv preprint arXiv:2301.04246* (2023).
[18] Chee, F. Y. and Mukherjee, S. *Exclusive: ChatGPT in spotlight as EU's Breton bats for tougher AI rules*. Reuters, City, 2023.
[19] Smith, B. *Meeting the AI moment: advancing the future through responsible AI*. City, 2023.
[20] Lieu, T. *I'm a Congressman Who Codes. A.I. Freaks Me Out.*, City, 2023.
[21] *An Act drafted with the help of ChatGPT to regulate generative artificial intelligence models like ChatGPT.*, City, 2023.
[22] Helberger, N. and Diakopoulos, N. ChatGPT and the AI Act. *Internet Policy Review*, 12, 1 (2023).
[23] Veale, M. and Borgesius, F. Z. Demystifying the Draft EU Artificial Intelligence Act—Analysing the good, the bad, and the unclear elements of the proposed approach. *Computer Law Review International*, 22, 4 (2021), 97-112.
[24] Hacker, P. The European AI Liability Directives - Critique of a Half-Hearted Approach and Lessons for the Future. *Working Paper, https://arxiv.org/abs/2211.13960* (2022).
[25] Hacker, P. A legal framework for AI training data—from first principles to the Artificial Intelligence Act. *Law, Innovation and Technology*, 13, 2 (2021), 257-301.
[26] Fredrikson, M., Jha, S. and Ristenpart, T. *Model inversion attacks that exploit confidence information and basic countermeasures*. City, 2015.
[27] Veale, M., Binns, R. and Edwards, L. Algorithms that remember: model inversion attacks and data protection law. *Philosophical Transactions of the Royal Society A: Mathematical, Physical and Engineering Sciences*, 376, 2133 (2018), 20180083.
[28] Carlini, N., Hayes, J., Nasr, M., Jagielski, M., Sehwag, V., Tramèr, F., Balle, B., Ippolito, D. and Wallace, E. Extracting Training Data from Diffusion Models. *arXiv preprint arXiv:2301.13188* (2023).
[29] Douek, E. Content Moderation as Systems Thinking. *Harv. L. Rev.*, 136 (2022), 526.
[30] De Gregorio, G. Democratising online content moderation: A constitutional framework. *Computer Law & Security Review*, 36 (2020), 105374.
[31] Heldt, A. P. *EU Digital Services Act: The white hope of intermediary regulation*. Palgrave, City, 2022.
[32] Meyer, P. *ChatGPT: How Does It Work Internally?*, City, 2022.
[33] Eifert, M., Metzger, A., Schweitzer, H. and Wagner, G. Taming the giants: The DMA/DSA package. *Common Market Law Review*, 58, 4 (2021), 987-1028.
[34] Laux, J., Wachter, S. and Mittelstadt, B. Taming the few: Platform regulation, independent audits, and the risks of capture created by the DMA and DSA. *Computer Law & Security Review*, 43 (2021), 105613.
[35] Kasy, M. and Abebe, R. *Fairness, equality, and power in algorithmic decision-making*. City, 2021.
[36] Barabas, C., Doyle, C., Rubinovitz, J. and Dinakar, K. *Studying up: reorienting the study of algorithmic fairness around issues of power*. City, 2020.
[37] Koops, E. *Should ICT Regulation Be Technology-Neutral?* TMC Asser Press, City, 2006.

[38] Bhuta, N., Beck, S. and Geiβ, R. *Autonomous weapons systems: law, ethics, policy*. Cambridge University Press, 2016.

[39] Sassoli, M. Autonomous weapons and international humanitarian law: Advantages, open technical questions and legal issues to be clarified. *International Law Studies*, 90, 1 (2014), 1.

[40] Bommasani, R., Hudson, D. A., Adeli, E., Altman, R., Arora, S., von Arx, S., Bernstein, M. S., Bohg, J., Bosselut, A. and Brunskill, E. On the opportunities and risks of foundation models. *arXiv preprint arXiv:2108.07258* (2021).

[41] Ganguli, D., Hernandez, D., Lovitt, L., Askell, A., Bai, Y., Chen, A., Conerly, T., Dassarma, N., Drain, D. and Elhage, N. Predictability and surprise in large generative models. *ACM Conference on Fairness, Accountability, and Transparency* (2022), 1747-1764.

[42] Hoffmann, J., Borgeaud, S., Mensch, A., Buchatskaya, E., Cai, T., Rutherford, E., Casas, D. d. L., Hendricks, L. A., Welbl, J. and Clark, A. Training compute-optimal large language models. *arXiv preprint arXiv:2203.15556* (2022).

[43] Vaswani, A., Shazeer, N., Parmar, N., Uszkoreit, J., Jones, L., Gomez, A. N., Kaiser, Ł. and Polosukhin, I. Attention is all you need. *Advances in neural information processing systems*, 30 (2017).

[44] Devlin, J., Chang, M.-W., Lee, K. and Toutanova, K. Bert: Pre-training of deep bidirectional transformers for language understanding. *arXiv preprint arXiv:1810.04805* (2018).

[45] Radford, A., Narasimhan, K., Salimans, T. and Sutskever, I. Improving language understanding by generative pre-training (2018).

[46] Lewis, M., Liu, Y., Goyal, N., Ghazvininejad, M., Mohamed, A., Levy, O., Stoyanov, V. and Zettlemoyer, L. Bart: Denoising sequence-to-sequence pre-training for natural language generation, translation, and comprehension. *arXiv preprint arXiv:1910.13461* (2019).

[47] Brown, T., Mann, B., Ryder, N., Subbiah, M., Kaplan, J. D., Dhariwal, P., Neelakantan, A., Shyam, P., Sastry, G. and Askell, A. Language models are few-shot learners. *Advances in neural information processing systems*, 33 (2020), 1877-1901.

[48] Kim, B., Kim, H., Lee, S.-W., Lee, G., Kwak, D., Jeon, D. H., Park, S., Kim, S., Kim, S. and Seo, D. What changes can large-scale language models bring? intensive study on hyperclova: Billions-scale korean generative pretrained transformers. *arXiv preprint arXiv:2109.04650* (2021).

[49] Bienert, J. and Klös, H.-P. *Große KI-Modelle als Basis für Forschung und wirtschaftliche Entwicklung*. IW-Kurzbericht, 2022.

[50] Radford, A., Kim, J. W., Hallacy, C., Ramesh, A., Goh, G., Agarwal, S., Sastry, G., Askell, A., Mishkin, P., Clark, J., Krueger, G. and Sutskever, I. Learning Transferable Visual Models From Natural Language Supervision. In *Proceedings of the Proceedings of the 38th International Conference on Machine Learning* (Proceedings of Machine Learning Research, 2021). PMLR, [insert City of Publication],[insert 2021 of Publication].

[51] Pham, H., Dai, Z., Ghiasi, G., Kawaguchi, K., Liu, H., Yu, A. W., Yu, J., Chen, Y.-T., Luong, M.-T. and Wu, Y. Combined scaling for open-vocabulary image classification. *arXiv e-prints* (2021), arXiv: 2111.10050.

[52] Dao, T., Fu, D. Y., Ermon, S., Rudra, A. and Ré, C. Flashattention: Fast and memory-efficient exact attention with io-awareness. *arXiv preprint arXiv:2205.14135* (2022).

[53] Geiping, J. and Goldstein, T. Cramming: Training a Language Model on a Single GPU in One Day. *arXiv preprint arXiv:2212.14034* (2022).

[54] OECD *Measuring the Environmental Impacts of AI Compute and Applications: The AI Footprint*. City, 2022.

[55] Freitag, C., Berners-Lee, M., Widdicks, K., Knowles, B., Blair, G. S. and Friday, A. The real climate and transformative impact of ICT: A critique of estimates, trends, and regulations. *Patterns*, 2, 9 (2021), 100340.

[56] ACM, T. P. C. *ACM TechBrief: Computing and Climate Change*. City, 2021.

[57] Cowls, J., Tsamados, A., Taddeo, M. and Floridi, L. The AI gambit: leveraging artificial intelligence to combat climate change—opportunities, challenges, and recommendations. *AI & Society* (2021), 1-25.

[58] Taddeo, M., Tsamados, A., Cowls, J. and Floridi, L. Artificial intelligence and the climate emergency: Opportunities, challenges, and recommendations. *One Earth*, 4, 6 (2021), 776-779.

[59] Ananthaswamy, A. *The Physics Principle That Inspired Modern AI Art*. City, 2023.

[60] Sohl-Dickstein, J., Weiss, E., Maheswaranathan, N. and Ganguli, S. *Deep unsupervised learning using nonequilibrium thermodynamics*. PMLR, City, 2015.

[61] Liu, P., Yuan, W., Fu, J., Jiang, Z., Hayashi, H. and Neubig, G. Pre-train, Prompt, and Predict: A Systematic Survey of Prompting Methods in Natural Language Processing. *ACM Computing Surveys*, 55 (2021), 1 - 35.

[62] Ouyang, L., Wu, J., Jiang, X., Almeida, D., Wainwright, C. L., Mishkin, P., Zhang, C., Agarwal, S., Slama, K. and Ray, A. Training language models to follow instructions with human feedback. *arXiv preprint arXiv:2203.02155* (2022).

[63] Luccioni, A. S. and Viviano, J. D. What's in the Box? A Preliminary Analysis of Undesirable Content in the Common Crawl Corpus. *arXiv preprint arXiv:2105.02732* (2021).
[64] Nadeem, M., Bethke, A. and Reddy, S. StereoSet: Measuring stereotypical bias in pretrained language models. *arXiv preprint arXiv:2004.09456* (2020).
[65] Zhao, Z., Wallace, E., Feng, S., Klein, D. and Singh, S. *Calibrate before use: Improving few-shot performance of language models*. PMLR, City, 2021.
[66] Bai, Y., Kadavath, S., Kundu, S., Askell, A., Kernion, J., Jones, A., Chen, A., Goldie, A., Mirhoseini, A. and McKinnon, C. Constitutional AI: Harmlessness from AI Feedback. *arXiv preprint arXiv:2212.08073* (2022).
[67] Perrigo, B. *OpenAI Used Kenyan Workers on Less Than $2 Per Hour to Make ChatGPT Less Toxic*. City, 2023.
[68] Bertuzzi, L. *Leading MEPs exclude general-purpose AI from high-risk categories – for now*. City, 2022.
[69] Bertuzzi, L. *AI Act: EU Parliament's crunch time on high-risk categorisation, prohibited practices*. City, 2023.
[70] Bertuzzi, L. *AI Act: MEPs close in on rules for general purpose AI, foundation models*. City, 2023.
[71] Bar-Ziv, S. and Elkin-Koren, N. Behind the scenes of online copyright enforcement: Empirical evidence on notice & takedown. *Conn. L. Rev.*, 50 (2018), 339.
[72] Cobia, J. The digital millennium copyright act takedown notice procedure: Misuses, abuses, and shortcomings of the process. *Minn. JL Sci. & Tech.*, 10 (2008), 387.
[73] Urban, J. M. and Quilter, L. Efficient process or chilling effects-takedown notices under Section 512 of the Digital Millennium Copyright Act. *Santa Clara Computer & High Tech. LJ*, 22 (2005), 621.
[74] Hacker, P., Cordes, J. and Rochon, J. Regulating Gatekeeper AI and Data: Transparency, Access, and Fairness under the DMA, the GDPR, and beyond. *Working Paper*, https://arxiv.org/abs/2212.04997 (2022).
[75] Morais Carvalho, J., Arga e Lima, F. and Farinha, M. Introduction to the Digital Services Act, Content Moderation and Consumer Protection. *Revista de Direito e Tecnologia*, 3, 1 (2021), 71-104.
[76] Riehm, T. and Meier, S. Product Liability in Germany. *J. Eur. Consumer & Mkt. L.*, 8 (2019), 161.
[77] Spindler, G. Die Vorschläge der EU-Kommission zu einer neuen Produkthaftung und zur Haftung von Herstellern und Betreibern Künstlicher Intelligenz. *Computer und Recht* (2022), 689-704.
[78] Wagner, G. Liability Rules for the Digital Age - Aiming for the Brussels Effect. *European Journal of Tort Law (forthcoming)* (2023), , https://ssrn.com/abstract=4320285.
[79] Bubeck, S., Chandrasekaran, V., Eldan, R., Gehrke, J., Horvitz, E., Kamar, E., Lee, P., Lee, Y. T., Li, Y. and Lundberg, S. Sparks of artificial general intelligence: Early experiments with GPT-4. *arXiv preprint arXiv:2303.12712* (2023).
[80] Goodfellow, I., Bengio, Y. and Courville, A. *Deep Learning*. MIT Press, 2016.
[81] Bennett, C. C. and Hauser, K. Artificial intelligence framework for simulating clinical decision-making: A Markov decision process approach. *Artificial intelligence in medicine*, 57, 1 (2013), 9-19.
[82] Geradin, D., Karanikioti, T. and Katsifis, D. GDPR Myopia: how a well-intended regulation ended up favouring large online platforms. *European Competition Journal*, 17, 1 (2021), 47-92.
[83] Bertuzzi, L. *MEPs seal the deal on Artificial Intelligence Act*. City, 2023.
[84] Liang, P., Bommasani, R., Creel, K. and Reich, R. *The time is now to develop community norms for the release of foundation models*. City, 2022.
[85] Bornstein, M., Appenzeller, G. and Casado, M. *Who Owns the Generative AI Platform?* , City, 2023.
[86] Blair, R. D. and Kaserman, D. L. *Law and economics of vertical integration and control*. Academic Press, 2014.
[87] Perry, M. K. Vertical integration: Determinants and effects. *Handbook of industrial organization*, 1 (1989), 183-255.
[88] Kolasky, W. J. and Dick, A. R. The merger guidelines and the integration of efficiencies into antitrust review of horizontal mergers. *Antitrust LJ*, 71 (2003), 207.
[89] Stuyck, J. *Consumer Concepts in EU Secondary Law*. De Gruyter, City, 2015.
[90] Micklitz, H.-W., Stuyck, J., Terryn, E. and School, I. C. *Consumer law*. Hart London, 2010.
[91] Blaschke, T. and Bajorath, J. Fine-tuning of a generative neural network for designing multi-target compounds. *Journal of Computer-Aided Molecular Design*, 36, 5 (2022/05/01 2022), 363-371.
[92] Ziegler, D. M., Stiennon, N., Wu, J., Brown, T. B., Radford, A., Amodei, D., Christiano, P. and Irving, G. Fine-tuning language models from human preferences. *arXiv preprint arXiv:1909.08593* (2019).
[93] Widder, D. G. and Nafus, D. Dislocated Accountabilities in the AI Supply Chain: Modularity and Developers' Notions of Responsibility. *arXiv preprint arXiv:2209.09780* (2022).
[94] Meyers, J. M. Artificial intelligence and trade secrets. *Landslide*, 11 (2018), 17.
[95] Bertuzzi, L. *Leading EU lawmakers propose obligations for General Purpose AI*. City, 2023.
[96] Drexl, J., Hilty, R., Desaunettes-Barbero, L., Globocnik, J., Gonzalez Otero, B., Hoffmann, J., Kim, D., Kulhari, S., Richter, H. and Scheuerer, S. Artificial Intelligence and Intellectual Property Law-Position Statement of the Max Planck

Institute for Innovation and Competition of 9 April 2021 on the Current Debate. *Max Planck Institute for Innovation & Competition Research Paper*, 21-10 (2021).
[97] Calvin, N. and Leung, J. Who owns artificial intelligence? A preliminary analysis of corporate intellectual property strategies and why they matter. *Future of Humanity Institute, February* (2020).
[98] Deeks, A. The judicial demand for explainable artificial intelligence. *Columbia Law Review*, 119, 7 (2019), 1829-1850.
[99] Hacker, P. and Passoth, J.-H. Varieties of AI Explanations under the Law. From the GDPR to the AIA, and Beyond. *International Conference on Extending Explainable AI Beyond Deep Models and Classifiers* (2022), 343-373.
[100] Schmidt-Wudy, F. *Art. 15 DSGVO*. City, 2023.
[101] McKown, J. R. Discovery of Trade Secrets. *Santa Clara Computer & High Tech. LJ*, 10 (1994), 35.
[102] Roberts, J. Too little, too late: Ineffective assistance of counsel, the duty to investigate, and pretrial discovery in criminal cases. *Fordham Urb. LJ*, 31 (2003), 1097.
[103] Shepherd, G. B. An empirical study of the economics of pretrial discovery. *International Review of Law and Economics*, 19, 2 (1999), 245-263.
[104] Subrin, S. N. Discovery in Global Perspective: Are We Nuts. *DePaul L. Rev.*, 52 (2002), 299.
[105] Kötz, H. Civil justice systems in Europe and the United States. *Duke J. Comp. & Int'l L.*, 13 (2003), 61.
[106] Daniel, P. F. Protecting Trade Secrets from Discovery. *Tort & Ins. LJ*, 30 (1994), 1033.
[107] Shavell, S. *Foundations of Economic Analysis of Law*. Harvard U Press, 2004.
[108] Shavell, S. On liability and insurance. *Bell Journal of Economics*, 13 (1982), 120-132.
[109] Adams-Prassl, J., Binns, R. and Kelly-Lyth, A. Directly Discriminatory Algorithms. *The Modern Law Review* (2022).
[110] Wachter, S. The Theory of Artificial Immutability: Protecting Algorithmic Groups Under Anti-Discrimination Law. *arXiv preprint arXiv:2205.01166* (2022).
[111] Wachter, S., Mittelstadt, B. and Russell, C. Why fairness cannot be automated: Bridging the gap between EU non-discrimination law and AI. *Computer Law & Security Review*, 41 (2021), 105567.
[112] Hacker, P. Teaching fairness to artificial intelligence: existing and novel strategies against algorithmic discrimination under EU law. *Common Market Law Review*, 55, 4 (2018), 1143-1186.
[113] Barocas, S. and Selbst, A. D. Big data's disparate impact. *California Law Review* (2016), 671-732.
[114] Wachter, S. Affinity profiling and discrimination by association in online behavioral advertising. *Berkeley Tech. LJ*, 35 (2020), 367.
[115] Hacker, P. UberPop, UberBlack, and the regulation of digital platforms after the Asociación Profesional Elite Taxi Judgment of the CJEU. *European review of contract law*, 14, 1 (2018), 80-96.
[116] Goodfellow, I., Pouget-Abadie, J., Mirza, M., Xu, B., Warde-Farley, D., Ozair, S., Courville, A. and Bengio, Y. Generative adversarial networks. *Communications of the ACM*, 63, 11 (2020), 139-144.
[117] Geiger, C., Frosio, G. and Bulayenko, O. The exception for Text and Data Mining (TDM) in the Proposed Directive on Copyright in the Digital Single Market-legal aspects. *Centre for International Intellectual Property Studies (CEIPI) Research Paper*, 2018-02 (2018).
[118] Rosati, E. The exception for text and data mining (TDM) in the proposed Directive on Copyright in the Digital Single Market: technical aspects. *European Parliament* (2018).
[119] Mourby, M., Cathaoir, K. Ó. and Collin, C. B. Transparency of machine-learning in healthcare: The GDPR & European health law. *Computer Law & Security Review*, 43 (2021), 105611.
[120] Zuiderveen Borgesius, F. J., Kruikemeier, S., Boerman, S. C. and Helberger, N. Tracking walls, take-it-or-leave-it choices, the GDPR, and the ePrivacy regulation. *Eur. Data Prot. L. Rev.*, 3 (2017), 353.
[121] Gruschka, N., Mavroeidis, V., Vishi, K. and Jensen, M. *Privacy issues and data protection in big data: a case study analysis under GDPR*. IEEE, City, 2018.
[122] Forgó, N., Hänold, S. and Schütze, B. The principle of purpose limitation and big data. *New technology, big data and the law* (2017), 17-42.
[123] Zarsky, T. Z. Incompatible: The GDPR in the age of big data. *Seton Hall L. Rev.*, 47 (2016), 995.
[124] Mondschein, C. F. and Monda, C. The EU's General Data Protection Regulation (GDPR) in a research context. *Fundamentals of clinical data science* (2019), 55-71.
[125] Peloquin, D., DiMaio, M., Bierer, B. and Barnes, M. Disruptive and avoidable: GDPR challenges to secondary research uses of data. *European Journal of Human Genetics*, 28, 6 (2020), 697-705.
[126] Staunton, C., Slokenberga, S. and Mascalzoni, D. The GDPR and the research exemption: considerations on the necessary safeguards for research biobanks. *European Journal of Human Genetics*, 27, 8 (2019), 1159-1167.

[127] Gil González, E. and De Hert, P. *Understanding the legal provisions that allow processing and profiling of personal data—an analysis of GDPR provisions and principles*. Springer, City, 2019.
[128] Donnelly, M. and McDonagh, M. Health research, consent and the GDPR exemption. *European journal of health law*, 26, 2 (2019), 97-119.
[129] Jones, M. L. and Kaminski, M. E. An American's Guide to the GDPR. *Denv. L. Rev.*, 98 (2020), 93.
[130] Hildebrandt, M. *Law for computer scientists and other folk*. Oxford University Press, 2020.
[131] Bonatti, P. A. and Kirrane, S. *Big Data and Analytics in the Age of the GDPR*. IEEE, City, 2019.
[132] Butterworth, M. The ICO and artificial intelligence: The role of fairness in the GDPR framework. *Computer Law & Security Review*, 34, 2 (2018), 257-268.
[133] Information), D. H. B. f. D. u. I. T. H. C. f. D. P. a. F. o. *Hessischer Datenschutzbeauftragter fordert Antworten zu ChatGPT*. City, 2023.
[134] Baden-Württemberg), D. L. f. d. D. u. d. I. B.-W. T. S. C. f. D. P. a. F. o. I. o. LfDI informiert sich bei OpenAI, wie ChatGPT datenschutzrechtlich funktioniert (24 April 2023 2023).
[135] Information), T. L. f. d. D. u. d. I. T. S. C. f. D. P. a. F. o. ChatGPT - auf dem Prüfstand (14 April 2023 2023).
[136] Reuters *French privacy watchdog investigating complaints about ChatGPT*. City, 2023.
[137] Swaton, C. ChatGPT sparks data protection debate in Europe (14 April 2023 2023).
[138] Board, E. D. P. *EDPB resolves dispute on transfers by Meta and creates task force on Chat GPT*. City, 2023.
[139] Personali, G. p. l. P. d. D. *Provvedimento del 30 marzo 2023 [9870832]* City, 2023.
[140] Satariano, A. *ChatGPT Is Banned in Italy as Regulators Cite Privacy Concerns*. City, 2023.
[141] Party, A. D. P. W. *Guidelines on Transparency under Regulation 2016/679, WP260 rev.01*. City, 2018.
[142] Personali, G. p. l. P. d. D. *Provvedimento dell'11 aprile 2023 [9874702]* City, 2023.
[143] Personali, G. p. l. P. d. D. *ChatGPT: OpenAI riapre la piattaforma in Italia garantendo più trasparenza e più diritti a utenti e non utenti europei / ChatGPT: OpenAI reinstates service in Italy with enhanced transparency and rights for european users and non-users*. City, 2023.
[144] Brundage, M., Avin, S., Clark, J., Toner, H., Eckersley, P., Garfinkel, B., Dafoe, A., Scharre, P., Zeitzoff, T. and Filar, B. The malicious use of artificial intelligence: Forecasting, prevention, and mitigation. *arXiv preprint arXiv:1802.07228* (2018).
[145] Kiela, D., Firooz, H., Mohan, A., Goswami, V., Singh, A., Fitzpatrick, C. A., Bull, P., Lipstein, G., Nelli, T. and Zhu, R. *The hateful memes challenge: Competition report*. PMLR, City, 2021.
[146] Kiela, D., Firooz, H., Mohan, A., Goswami, V., Singh, A., Ringshia, P. and Testuggine, D. The hateful memes challenge: Detecting hate speech in multimodal memes. *Advances in Neural Information Processing Systems*, 33 (2020), 2611-2624.
[147] Zellers, R., Holtzman, A., Rashkin, H., Bisk, Y., Farhadi, A., Roesner, F. and Choi, Y. Defending against neural fake news. *Advances in neural information processing systems*, 32 (2019).
[148] Seeha, S. *Prompt Engineering and Zero-Shot/Few-Shot Learning [Guide]*. City, 2022.
[149] Deb, M., Deiseroth, B., Weinbach, S., Schramowski, P. and Kersting, K. AtMan: Understanding Transformer Predictions Through Memory Efficient Attention Manipulation. *arXiv preprint arXiv:2301.08110* (2023).
[150] Heikkilä, M. *The viral AI avatar app Lensa undressed me—without my consent*. City, 2022.
[151] Bertuzzi, L. and Killeen, M. *Tech Brief: Meta's advertising business, next six month's agenda*. City, 2023.
[152] OECD. *Disinformation and Russia's war of aggression against Ukraine: Threats and governance responses*. 2022.
[153] Balakrishnan, V., Zhen, N. W., Chong, S. M., Han, G. J. and Lee, T. J. Infodemic and fake news–A comprehensive overview of its global magnitude during the COVID-19 pandemic in 2021: A scoping review. *International Journal of Disaster Risk Reduction* (2022), 103144.
[154] European, C., Directorate-General for Communications Networks, C., Technology, Hoboken, J., Quintais, J., Poort, J. and Eijk, N. *Hosting intermediary services and illegal content online : an analysis of the scope of article 14 ECD in light of developments in the online service landscape : final report*. Publications Office, 2019.
[155] Brüggemeier, G., Ciacchi, A. C. and O'Callaghan, P. *Personality rights in european tort law*. cambridge university press, 2010.
[156] Wilman, F. The Digital Services Act (DSA)-An Ooverview. *Available at SSRN 4304586* (2022).
[157] Gerdemann, S. and Spindler, G. Das Gesetz über digitale Dienste (Digital Services Act) (Part 2). *Gewerblicher Rechtsschutz und Urheberrecht* (2023), 115-125.
[158] Korenhof, P. and Koops, B.-J. Gender Identity and Privacy: Could a Right to Be Forgotten Help Andrew Agnes Online? *Working Paper, https://ssrn.com/abstract=2304190* (2014).

[159] Lianos, I. and Motchenkova, E. Market dominance and search quality in the search engine market. *Journal of Competition Law & Economics*, 9, 2 (2013), 419-455.
[160] Geroski, P. A. and Pomroy, R. Innovation and the evolution of market structure. *The journal of industrial economics* (1990), 299-314.
[161] Patterson, D., Gonzalez, J., Le, Q., Liang, C., Munguia, L.-M., Rothchild, D., So, D., Texier, M. and Dean, J. Carbon emissions and large neural network training. *arXiv preprint arXiv:2104.10350* (2021).
[162] Bertuzzi, L. *AI Act: MEPs want fundamental rights assessments, obligations for high-risk users*. City, 2023.
[163] Grinbaum, A. and Adomaitis, L. The Ethical Need for Watermarks in Machine-Generated Language. *arXiv preprint arXiv:2209.03118* (2022).
[164] Kirchenbauer, J., Geiping, J., Wen, Y., Katz, J., Miers, I. and Goldstein, T. A Watermark for Large Language Models. *arXiv preprint arXiv:2301.10226* (2023).
[165] Mitchell, E., Lee, Y., Khazatsky, A., Manning, C. D. and Finn, C. DetectGPT: Zero-Shot Machine-Generated Text Detection using Probability Curvature. *arXiv preprint arXiv:2301.11305* (2023).
[166] Solaiman, I. The Gradient of Generative AI Release: Methods and Considerations. *arXiv preprint arXiv:2302.04844* (2023).
[167] Solaiman, I., Brundage, M., Clark, J., Askell, A., Herbert-Voss, A., Wu, J., Radford, A., Krueger, G., Kim, J. W. and Kreps, S. Release strategies and the social impacts of language models. *arXiv preprint arXiv:1908.09203* (2019).
[168] Crootof, R. Artificial intelligence research needs responsible publication norms. *Lawfare Blog* (2019).
[169] Hoffmann-Riem, W. *Innovation und Recht-Recht und Innovation: Recht im Ensemble seiner Kontexte*. Mohr Siebeck, 2016.
[170] Bennett Moses, L. Regulating in the face of sociotechnical change (2016).
[171] Bennett Moses, L. Recurring dilemmas: The law's race to keep up with technological change. *U. Ill. JL Tech. & Pol'y* (2007), 239.

**APPENDIX H1: PROMPTS**

Prompt 1: What are large generative AI models?

Prompt 2: What distinguishes large generative AI models from other AI systems?

Prompt 3: Can you explain the technical foundations of large generative models in simple terms, so that an inexperienced reader understands it?

Prompt 4: What are the objectives, what are the obstacles when it comes to content moderation within large generative AI models?

Prompt 5: How does content moderation work at ChatGPT?